\documentclass[pre,twocolumn,nofootinbib,twoside,showpacs,superscriptaddress]{revtex4-1}
\usepackage{graphicx,amssymb,amsmath,epsfig}
\epsfclipon

\usepackage{xspace}
\usepackage{color}

\begin{document}
\title{The ordered phase of $O(N)$ model within the Non-Perturbative 
Renormalization Group}

\author{Marcela Pel\'aez and  Nicol\'as Wschebor}
\affiliation{Instituto de F\'{\i}sica, Facultad de Ingenier\'{\i}a, Universidad 
de la Rep\'ublica, 
J.H.y Reissig 565, 11000 Montevideo, Uruguay}

\begin{abstract}
In the present article we analyze Non-Perturbative Renormalization Group flow 
equations in the order phase of
$\mathbb{Z}_2$ and $O(N)$ invariant scalar models in the derivative expansion 
approximation scheme. We first
address the behavior of the leading order approximation (LPA), discussing for 
which regulators the flow is smooth and gives a convex free energy
and when it becomes singular. We improve the exact known solutions in the 
``internal'' region of the potential and exploit this solution
in order to implement a numerical algorithm that is much more stable that 
previous ones for $N>1$. After that, we study the flow equations
at second order of the Derivative Expansion and analyze how and when LPA results 
change. We also discuss the evolution of field renormalization
factors.
\end{abstract}
\date{\today}
\pacs{ }
\maketitle

\section{Introduction}

The low temperature regime of typical statistical models are notoriously more 
difficult to handle than the high temperature one.
In such case it is usual that coexistence between two different phases dominates 
the free energy and this property is not easily handled
by mean-field methods or perturbative expansions. This manifests itself as
violations of various exact thermodynamics properties, such as
the convexity of the free energy in terms of the order parameter. Convexity can 
be restored by hand in mean-field treatments via the
Maxwell construction but it is unclear how this strategy can be followed 
systematically when a perturbative expansion is performed around
the mean field. The reason is that the convexity property in the coexistence 
region corresponds to large fluctuations in the configuration
space and not to small ones around a mean field analysis.

Since the 90s a method has been developed (the so-called Non-Perturbative 
Renormalization Group 
\cite{Wetterich93,Ellwanger93,Tetradis94,Morris94,Morris94c}, NPRG),
that can easily handle the convexity
properties of the free energy. It was quickly understood 
\cite{Ringwald89,Tetradis92,Horikoshi98,Alexandre98,Kapoyannis00,Caillol12,
Zappala12}
that one of the simplest approximations implemented in such
scheme  naturally yields a convex free-energy. This approximation scheme, 
called Local Potential Approximation (LPA) is also able to address successfully a large variety of statistical problems at 
equilibrium, out of equilibrium
(for reviews on the subject see for example 
\cite{Berges02,Delamotte07,Canet11}), or even in more
difficult contexts, like in presence of quenched disorder (see for example 
\cite{Tissier11}).
Moreover, this approximation scheme can be seen as the leading order of
a systematic expansion of vertex functions in wave-numbers called the Derivative 
Expansion (DE) \cite{Golner86,Berges02,Delamotte07}. The DE has shown to be
extremely successful \cite{Berges02,Delamotte07}. It has been pushed in the case 
of a single scalar Ginzburg-Landau model up to order
$\mathcal{O}(\partial^6)$ \cite{Canet03,Chate12prox},
obtaining at that order results with a better precision than Borel-resumed 
six-order perturbative expansion results for
critical exponents.

Given these previous results it is natural to extend the LPA results to the 
low-temperature phase of the Ginzbug-Landau model beyond the LPA
to other models as those with $O(N)$ 
symmetry \cite{Zappala12}. One of the difficulty is that although the LPA preserves the convexity of the free energy,
its numerical implementation is generically unstable: at sufficiently small RG scales, the flow blows up. 
Even if
the convexity shows up, the free energy approaches singular points of the flow equations and it is difficult to
push the solution numerically to low momentum scales. This is in radical 
contrast with what happens in the high temperature phase or around a critical point. Given this problem, 
sophisticated numerical methods have been employed in order
to solve the LPA equation at low renormalization group scales 
\cite{Pangon09,Pangon10,Caillol12}. The difficulty with such approaches is that 
they are difficult to implement at next-to-leading orders of DE and even more difficult 
when applied to more sophisticated approximation schemes such as the Blaizot-M\'endez-Wschebor scheme
\cite{Blaizot05,Benitez09,Benitez12}.

In the present article, we extend previous studies of the low temperature phase 
of Ginzburg-Landau models within the NPRG. First, we
improve previous results in LPA approximation to $O(N)$ model, analyzing in 
detail the dependence in the regulator and number of fields. We show
that, contrarily to what could be expected, the numerical behavior of flow 
equations is much more stable when $N>1$ than for $N=1$. In both cases
we improve and exploit analytical results for the free-energy in the coexistence 
region in order to implement a simple numerical
algorithm that permits to explore much smaller renormalization group momentum 
scales. Then, we extend the numerical results
for $O(N)$ models to the DE at next-to-leading order.

\section{Non-perturbative Renormalization Group and the Derivative Expansion}

Before considering the behavior of the effective action in the low temperature 
phase, let us recall briefly the origin and uses
of NPRG equations. We present this formalism for a generic euclidean field 
theory with $N$ scalars fields $\varphi_i$, 
denoted collectively by $\varphi$, with Hamiltonian $H[\varphi]$. Then, we 
specialize to the case where $H[\varphi]$ is $O(N)$ or $\mathbb{Z}_2$ symmetric.
The NPRG equations, intimately related to Wilsonian Renormalization Group 
equations, connect the Hamiltonian to the full
Gibbs free energy (generating functional of 1-PI vertex functions). This 
relation is obtained by controlling the magnitude of
long wavelength field fluctuations with the help of an infra-red cut-off, which 
is implemented
\cite{Tetradis94,Ellwanger94a,Morris94b,Morris94c} by adding to the
Hamiltonian $H[\varphi]$ a regulator of the form
\begin{equation}
  \Delta H_k[\varphi] =\frac{1}{2} \int \frac{d^dq}{(2\pi)^d}
(R_k(q))_{ij} \varphi_i(q)\varphi_j(-q),
\end{equation}\normalsize
where $(R_k(q))_{ij}$ denotes a family of $k$-dependent family of ``cut-off functions''
to be specified below. Above and below, sums are understood for repeated 
internal indices. 
The role of $\Delta H_k$ is to suppress the fluctuations of $\varphi(q))$ with momenta $q 
\lesssim k$,
while leaving unaffected the modes with $q \gg k$. Accordingly,
typically $ R_{k}(q)_{ij} \sim k^2 \delta_{ij}$ when $ q \ll k$, 
and $R_{k}(q)_{ij}\to 0$ quickly when $ q\gtrsim k$.

One can define an effective Gibbs free energy corresponding to 
$H[\varphi]+\Delta H_k[\varphi]$ denoted by
$\Gamma_k[\phi]$, where $\phi$ is the average field, $\phi_i(x)=\left\langle 
\varphi_i(x)\right\rangle$ in presence of external sources.
When $k=\Lambda$, with $\Lambda$ the microscopic scale of the problem, all 
fluctuations are suppressed and $\Gamma_\Lambda[\phi]$ coincides
 with the Hamiltonian. As $k$ is lowered, more and more fluctuations are taken into 
account and when  $k\to 0$, all fluctuations are included
 and $\Gamma_{k=0}[\phi]$
 becomes the Gibbs free energy $\Gamma[\phi]$ (see e.g. \cite{Berges02}). The 
flow of $\Gamma_k[\phi]$ with $k$ is given by the Wetterich
 equation \cite{Tetradis94,Ellwanger94a,Morris94b,Morris94c}:
\begin{equation}
\label{NPRGeq}
\partial_k\Gamma_\kappa[\phi]=\frac{1}{2}\int \frac{d^dq}{(2\pi)^d} \mathrm{tr} 
\bigg\{ \partial_\kappa R_\kappa(q^2)
\left[\Gamma_k^{(2)}+R_k\right]^{-1}_{q;-q}\bigg\},
\end{equation}
where $\Gamma_k^{(2)}$ denotes the matrix of second derivatives of $\Gamma_k$ 
with respect to $\phi$ and
the trace is taken over internal indices.

From now on, we shall specialize to $O(N)$-symmetric models.
Since we are interested in the following in non-universal properties such as the free-energy for $T<T_c$,
we need in principle to consider general O(N)-invariant hamiltonians. NPRG equations have no 
difficulties to handle non-renormalizable
Hamiltonians and can even include a realistic microscopic structure of a given 
system like a specific lattice model in order to
analyse non universal properties \cite{Machado10,Rancon11}. However, for the 
purposes of the present article it is enough to
choose a simple $\varphi^4$ Ginzburg-Landau Hamiltonian given by
\begin{align}\label{HG-L}
H[\varphi] = \int d^{d}x\,&\left\lbrace{
\frac{1}{2}} \nabla \varphi_i(x)\cdot \nabla
\varphi_i(x) + \frac{r}{2} \,  \varphi_i(x) \varphi_i(x)\right.\nonumber\\
&\left. +
\frac{u}{4!} \,\left( \varphi_i(x) \varphi_i(x) \right)^2 \right\rbrace.
\end{align}
In order to preserve the $O(N)$ symmetry all along the flow, it is mandatory to 
consider a regulator respecting this symmetry.
This implies the use of a regulator of the form
\[
(R_\kappa(q))_{ij} \equiv R_\kappa(q) \delta_{ij} .
\]
In practice, we 
chose functions $R_k(q)$ of the
two types more frequently used in the literature.
The first one corresponds to the $\theta$-regulator \cite{Litim} equal, up to 
field renormalizations to
\begin{equation}
\label{thetareg}
R_k(q)= (k^2-q^2) \theta(1-q^2/k^2).
\end{equation}
The second one corresponds to infinitely differentiable regulators that decrease 
rapidly when $q \gg k$. In practice, for
numerical implementations, we use the standard exponential regulator that 
is, up to field renormalizations
\begin{equation}
\label{expreg}
 R_k(q)= \alpha \frac{q^2}{e^{q^2/k^2}-1}.
\end{equation}
Here the pre-factor $\alpha$ has been included in order to study typical 
regulator dependence of various results \cite{Canet02b}.

Before considering our specific analysis of the low temperature phase, let us 
discuss briefly the approximation scheme employed
in the present article, the DE. This corresponds to expanding the Gibbs free 
energy in the derivatives of the field while keeping any other possible field dependence. For example, at leading order 
(LPA), it corresponds to
taking an arbitrary effective potential and the bare form of the terms including derivatives of the field:
\begin{align}\label{LPA}
\Gamma_k[\phi] = \int d^{d}x\,\left\lbrace
\frac{1}{2} \nabla \phi_i(x)\cdot \nabla
\phi_i(x) + U_k(\rho)\right\rbrace,
\end{align}
where, $\rho=\phi_i(x) \phi_i(x)/2$. At next-to-leading order (also 
called $\mathcal{O}(\partial^2)$ order), all possible $O(N)$-invariant terms involving two derivatives must
be included in the {\it ansatz} of $\Gamma_k$:
\begin{align}
\label{ansatzD2}
\Gamma_k(\phi)=\int d^dx &\left\lbrace 
U_k(\rho)+\frac{1}{2}Z_k(\rho)\nabla\phi_i\cdot \nabla\phi_i\right. \nonumber\\
&+\left.\frac{1}{4}Y_k(\rho)\nabla\rho\cdot\nabla\rho\right\rbrace+\mathcal{O}
(\partial^4)
\end{align}
In the particular case of a single scalar field $N=1$, the third term is 
redundant and one can in this specific case take $Y=0$
without loss of generality. In that particular case, the DE has been pushed to 
order $\mathcal{O}(\partial^6)$ \cite{Chate12prox} for the study
of critical exponents.

\begin{figure}[ht]
\begin{center}
\includegraphics[width=8cm]{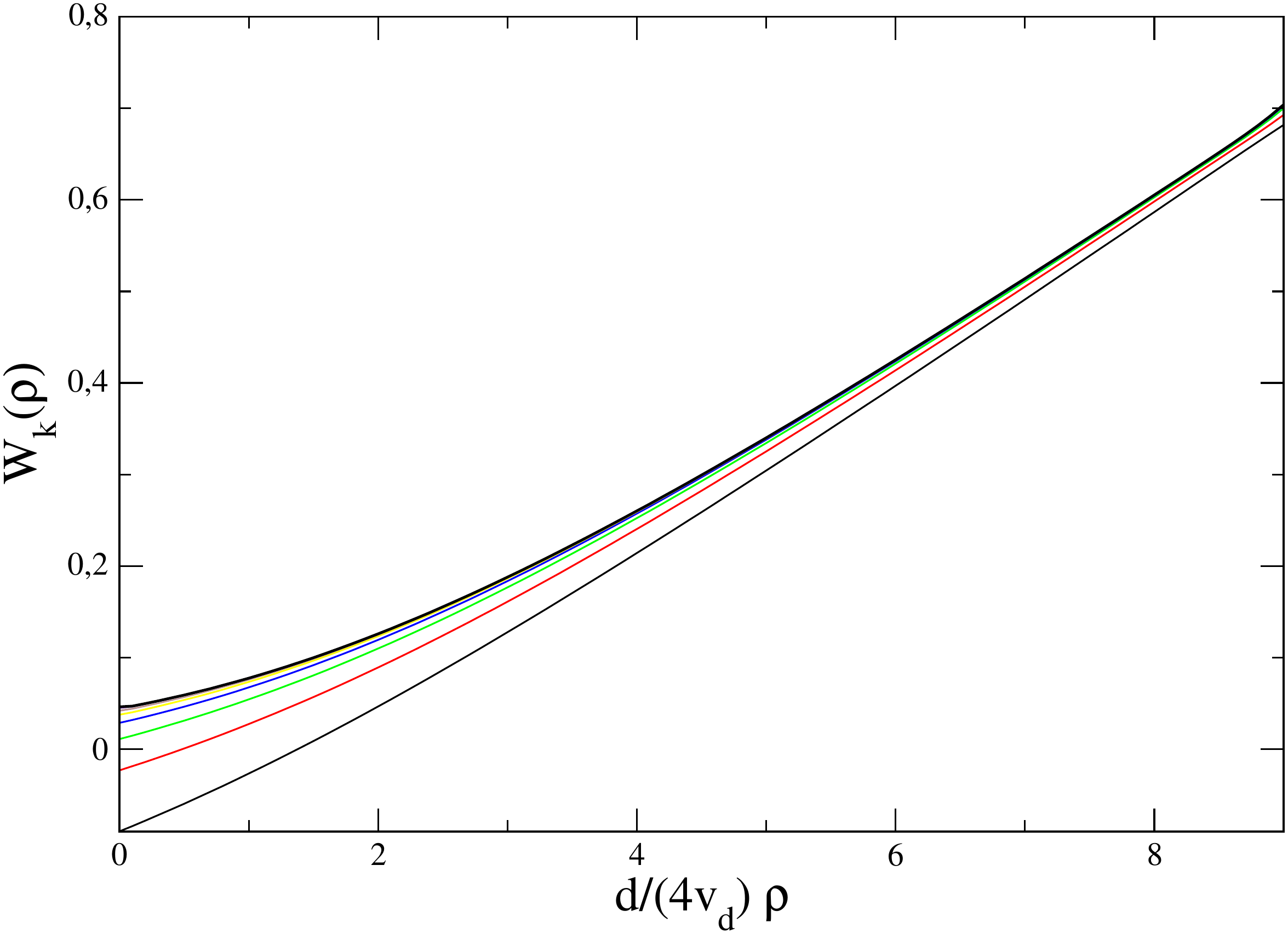}\\
\includegraphics[width=8cm]{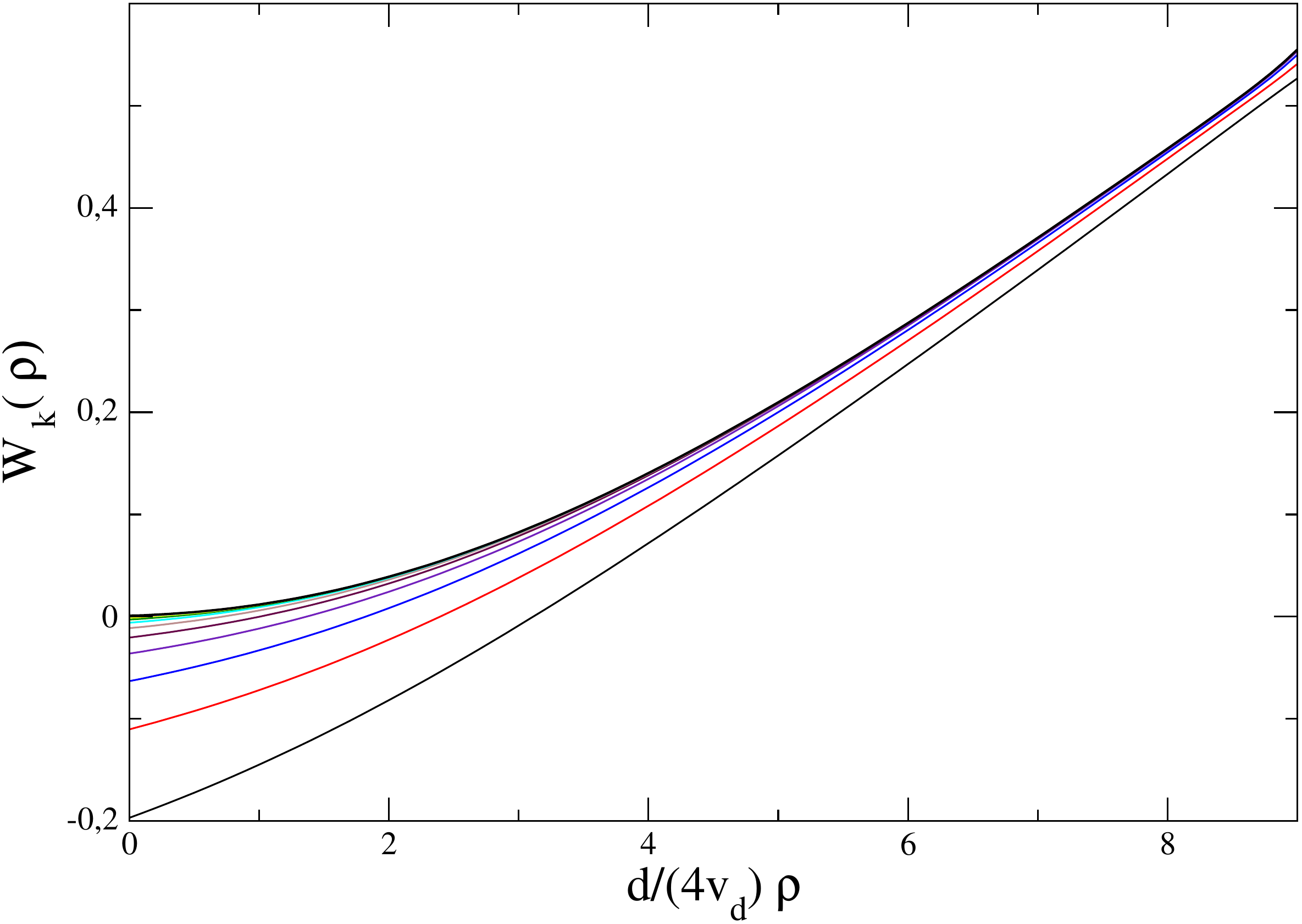}\\
\includegraphics[width=8cm]{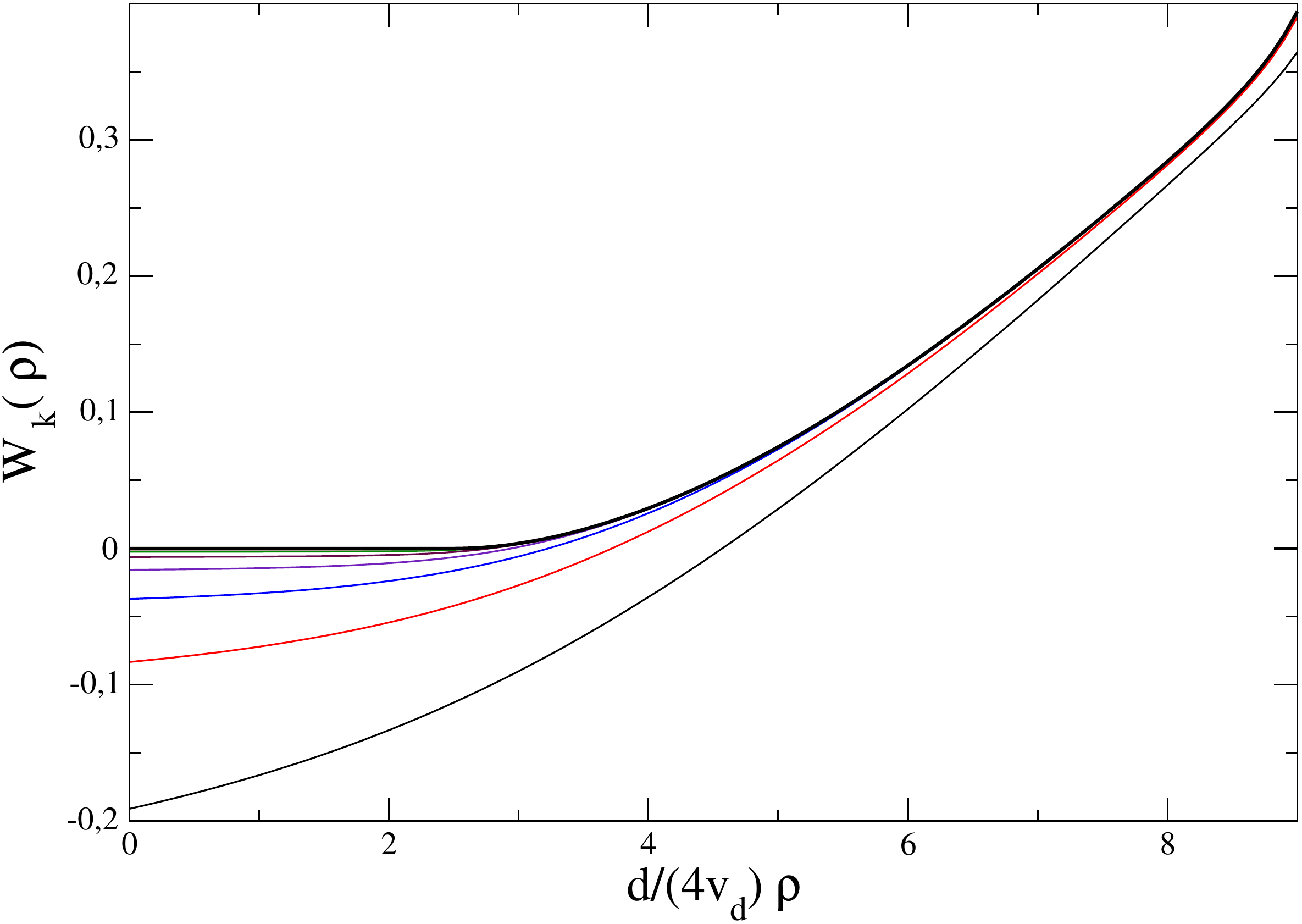}
\end{center}
\caption{\label{FRFS} (color online) Top: typical evolution of the potential to 
the high temperature phase.
Center: Typical evolution near the critical point.
Bottom: Typical evolution of the potential in the low temperature phase.}
\end{figure}

Finally, let us mention some difficulties encountered in previous works where the
low temperature phase of the $O(N)$ models has been studied with the DE.
We give a more 
detailed analysis of the corresponding equations in the following
sections. The difficulties appear already at LPA level.
the LPA. The flow equation of the derivative of the potential $W_k(\rho)=\partial_\rho 
U_k(\rho)$ reads
\begin{align}
\label{lpaeqthetareg}
\partial_tW_k &=-\frac{4v_d}{d}k^{d+2}\left(\frac{3W'_k+2\rho 
W''_k}{(k^2+W_k+2\rho W'_k)^2}+\frac{(N-1)W'_k}{(k^2+W_k)^2}\right)
\end{align}
where, $v_d^{-1}=2^{d+1}\pi^{d/2}\Gamma(d/2)$, $t=\log(k/\Lambda)$ and the $\theta$-regulator, Eq.~(\ref{thetareg}), has been used.
At the beginning of the flow, $U_\Lambda(\rho)$
is the bare potential Eq.~(\ref{HG-L}). Accordingly,
\begin{equation}\label{bareW}
 W_\Lambda(\rho)=r+\frac u 3 \rho.
\end{equation}
One can control in which phase the system is by computing the position of the minimum of 
the effective potential $U_{k=0}$ at $k=0$. At the mean-field level, the minimum
corresponds to $\rho_0$, the zero of $W_\Lambda(\rho)$, that is $-3 r/u$ if $r<0$ 
or zero if $r\geq 0$. Fluctuations tend to lower the value of
the average $\langle\varphi\rangle$ of the field and thus of the value of the running
minimum $\rho_{0}(k)$ of $U_k$ when $k$ is decreased. When $T>T_c$, the running minimum
hits the origin for a non-vanishing value of $k$: $\rho_{0}(k>0)=0$ while at $T_c$ it collapses
with the origin right at $k=0$. At fixed $u$, the value of $r$ for which the transition occurs
is therefore negative, $\rho_{0}(k>0)>0$ and $\rho_{0}(k=0)=0$. For $T<T_c$, $\rho_{0}(K)$
remains positive even for $k=0$ and thus, for the ``internal region of the potential'',
that is $\rho<\rho_{0}(k)$, $W_k(\rho)<0$, see Fig.~\ref{FRFS}. This is the origin of the
difficulties since poles in the denominator of the flow equation (\ref{lpaeqthetareg}) can appear becase of this negative sign if the regulator is not large enough.
For the $\theta$-regulator one must require in order
to avoid initial singularities that
\begin{equation}
 \Lambda^2+W_\Lambda>0\hspace{.5cm}\mathrm{and}\hspace{.5cm} 
\Lambda^2+W_\Lambda+2\rho W_\Lambda'>0
\end{equation}
that is $\Lambda^2+r>0$. The problem
is even worst: it has been shown, and will be discussed in detail in next 
sections, that when $k\to 0$ in the low temperature phase
the flow brings the potential to the regime where $0< W_k(\rho)+k^2 \ll k^2$ 
that is numerically even more demanding. Similar observations applies to the LPA 
equation
with other regulators \cite{Berges02}.

\

There are also good news, as has been analyzed before 
\cite{Tetradis92,Tetradis95}. The first one is that in LPA and for some 
regulators, the singularity works
as a barrier in flow equations and consequently the singularity is approached 
but never reached. Accordingly, the effective potential behaves
as $W_k(\rho)\sim -k^2$ in all the internal region.
This implies that in those cases, the
LPA approximation preserves the convexity of the physical free-energy, that 
becomes flat in the internal region for $k\to 0$, as is shown 
in the bottom figure of Fig.~\ref{FRFS}.
The second good news is that in the neighborhood of the singularity, the NPRG equation simplifies and
analytical solutions can be obtained in that regime \cite{Tetradis92}. In this 
article following a suggestion made in \cite{Berges02} we exploit
analytical solutions in the internal region of the potential in order to 
construct an
efficient algorithm for the broken phase. It must be stressed that in order to 
do so,
it has been necessary to improve considerably previous analytical results. In 
fact, previous results from \cite{Tetradis95} were only valid at small values of 
the fields but
in order to implement the numerical scheme just mentioned it is necessary to 
know the analytical form of the solution for large values of $\rho$.
Such solution is presented here for the LPA approximation and exploited in order 
to improve qualitatively the quality of the numerical treatment.

\section{Local Potential Approximation}
\label{secLPA}
The properties of NPRG equations in LPA approximation in the broken phase have 
already been analyzed in the literature both analytically and numerically
\cite{Tetradis92,Tetradis95,Berges02}.
In the present section we briefly review some of these works, generalize 
them to other cases and also show some limitations of previous results.
After that we exploit the analytical results in order to implement a simple 
numerical analysis that is
significantly more stable than previously considered \footnote{In 
Ref.~\cite{Caillol12} an efficient algorithm has been implemented for the LPA
approximation of the $N=1$ case. It is important
to observe, however, that it exploits many specificities of this particular case 
and that it is not trivial to generalize such procedure
for other values of $N$, or in more involved approximation schemes.}.

The NPRG equation for the derivative of the effective potential for a generic 
regulator profile $R_k(q)$ reads \cite{Berges02}:
\begin{align}
\label{lpaW}
\partial_tW_k=-\frac{1}{2}\int 
\frac{d^dq}{(2\pi)^d}&\partial_tR_k(q)\Big\{\frac{(N-1)W'_k}{(q^2+R_k(q)+W_k)^2}
\nonumber\\
&+\frac{3W'_k+2\rho W''_k}
{(q^2+R_k(q)+W_k+2\rho W'_k)^2}\Big\}
\end{align}
Generalizing the discussion of the introduction, if the flow avoids the presence 
of singularities, one must have for all $\rho,\, q$ and $k$,
\begin{equation}
\left\{
\begin{array}{l}
q^2+R_k(q)+W_k>0,\\
q^2+R_k(q)+W_k+2\rho W'_k>0
\end{array}
\right.
\end{equation}
Now, in the internal region of the potential, one must have for any $k$, 
$W_k<0$. Accordingly, given that $R_k(q)\sim \mathcal{O}(k^2)$,
one concludes that, for $q\lesssim k$, $W_k(\rho)=\mathcal{O}(k^2)$ (or smaller). On the 
other hand, in the low-temperature phase, the effective potential should
have a non trivial behavior in terms of the physical dimensionful field, or 
equivalently in terms of $\rho$ \footnote{It must be mentioned that 
when the system is near a critical regime, an hybrid procedure may be 
convenient. That is, one can take a re-scaling of the field that introduces
the standard dimensionless fields at values of $k$ much larger that the physical 
scales of the problem and becomes just a finite rescaling in the opposite case}.
This motivates the use of $w_k(\rho)=W_k(\rho)/k^2$ instead of $W_k(\rho)$. It is
convenient to introduce also the dimensionless function $r(y)$ defined by 
$R_k(q^2)=q^2 r(q^2/k^2)$. With these definitions, the equation
for $w_k(\rho)$ reads
\begin{align}
\label{eqwk}
&\partial_tw_k=-2 w_k+v_d k^{d-2}\int_0^\infty dy\,y^{d/2+1}\, r'(y)\nonumber\\
&\times\Big\{\frac{(N-1)w'_k}{(y(1+r(y))+w_k)^2}+\frac{3w'_k+2\rho w''_k}
{(y(1+r(y))+w_k+2\rho w'_k)^2}\Big\}
\end{align}
This is in contrast with the usual set of variables used in studies of the 
critical domain, where $w_k$ is studied as a function of the
{\it dimensionless field} which is, at LPA level, 
$\tilde\rho=\rho/k^{d-2}$. 

In the rest of this section we study this equation for various values of 
$N$ and for various regulators both analytically and numerically.
We show that the LPA equation does not avoid the existence of singularities 
of the flow unless a sufficiently strong regulator is included.
In particular, the $\theta$-regulator (\ref{thetareg}) does respect this 
property. Smooth regulators respect this property also if $1+R'(q=0)<0$ 
(corresponding
to the case $\alpha>2$ for exponential regulators (\ref{expreg})) \cite{Tetradis92,Tetradis95,Berges02}. When this 
property is not fulfilled, the flow brings the potential to the
singularity at $W_k+R(q)=0$ (typically at $\rho=0$ and $q=0$). This case was not 
fully addressed before in the literature even if such possibility
was suggested in \cite{Tetradis95,Berges02}.

\subsection{Large $N$}

We first analyze the large $N$ limit of Eq.~(\ref{eqwk}). This has been done 
long time ago \cite{Tetradis95} but we include it
here for completeness. Moreover in this case many calculations 
can be done analytically and this motivates
the general behavior of the potential obtained in the general case. 
The large $N$ limit is taken in the usual way (see, for example, 
\cite{D'Attanasio97}). It is simpler to analyze it for the dimensionful 
derivative of the
potential $W_k$. The coupling $u$ is of order $1/N$ and $V_k$
and $\rho$ are of order $N$. Accordingly, $W_k$ is of order $1$ and the 
large $N$ limit of Eq.~(\ref{lpaW}) is:
\begin{align}
\label{lpalargeN}
\partial_tW_k=-\frac{N}{2}W'_k \int 
\frac{d^dq}{(2\pi)^d}\frac{\partial_tR_k(q)}{(q^2+R_k(q)+W_k)^2}
\end{align}
An implicit solution of this differential equation can be obtained by 
considering the inverse function $\rho=F_k(W)$ \cite{Tetradis95}. It satisfies 
$F_k'(W)=1/W_k'(\rho)$ and $\partial_t 
F_k(W)=-F_k'(W)\partial_tW_k(\rho)$. Accordingly
\begin{equation}
\label{eqforFlargeN}
\partial_t F_k(W)=\frac{N}{2}\int 
\frac{d^dq}{(2\pi)^d}\frac{\partial_tR_k(q)}{(q^2+R_k(q)+W)^2}
\end{equation}
In this equation $W$ must be seen as an independent variable and consequently it 
can be integrated:
\begin{align}
\label{solFlargeN}
F_k(W)-F_\Lambda(W)=-\frac N 2\int 
&\frac{d^dq}{(2\pi)^d}\Big\{\frac{1}{q^2+R_k(q)+W}\nonumber\\
&-\frac{1}{q^2+R_{\Lambda}(q)+W}\Big\}
\end{align}
Given an initial condition for the potential, one can invert it in order to 
obtain $F_\Lambda(W)$. For example, for a Hamiltonian of
the form (\ref{HG-L}), one obtains by inverting the relation between $W_k(\rho)$ 
and $\rho$:
\begin{equation}
 F_\Lambda(W)=\frac 3 u (W-r)
\end{equation}
If $\Lambda$ much larger than any other physical scale, one can absorb 
for $d<4$ the dependence on $\Lambda$ in 
a renormalization of the parameter $r$, obtaining an implicit equation for 
$W_k(\rho)$:
\begin{align}
\label{sol3FlargeN}
\rho&-\frac 3 u (W_k(\rho)-\tilde{r})=-v_d N k^{d-2} \int_0^\infty 
dy\,y^{d/2-1}\nonumber\\
&\times\Big(\frac{1}{y(1+r(y))+W_k(\rho)/k^2}-\frac 1 y\Big).
\end{align}
where
\begin{equation}
\label{tilder}
 \tilde 
r=r+\frac{Nu}{6}\int\frac{d^dq}{(2\pi)^d}\Big\{\frac{1}{q^2}-\frac{1}{q^2+R_{
\Lambda}}\Big\}
\end{equation}
is the renormalized mass parameter. 
We can see here that the minimum of the potential goes to zero when $k\to 0$
only if $\tilde r=0$. One deduces that $\tilde r \propto (T-T_c)$ near the phase 
transition. We use this equation now in order to study the behavior for various 
regulators and, in particular,
analyze how the convexity is approached in the low temperature phase and if and 
when the singularity can be reached at a non zero value of $k$.
As expected, there is only a broken phase for $d>2$ because for $d\leq 2$ the 
integral in (\ref{tilder}) is infrared divergent.

Let us consider now how this equation behaves for specific regulators. Let us 
consider first the $\theta$-regulator (\ref{thetareg}), that allows
integrals to be done analytically at integer dimensions. For example, for $d=3$,
\begin{align}
\label{sol3FlargeNthetareg}
\Big(&\frac 3 u (w_k k^2 -\tilde{r})-\rho\Big)/(2 v_3 N k)=\nonumber\\
&\left\{
\begin{array}{ll}
-1+\frac 1 
{3+3w_k(\rho)}-\sqrt{w_k}\mathrm{arctan}(\sqrt{w_k})\;&\mathrm{if}\,w_k\geq 0\\
-1+\frac 1 
{3+3w_k}+\sqrt{|w_k|}\mathrm{arctanh}(\sqrt{|w_k|})\;&\mathrm{if}\,w_k< 0
\end{array}
\right.
\end{align}

\begin{figure}[ht]
\centering\includegraphics[width=8cm]{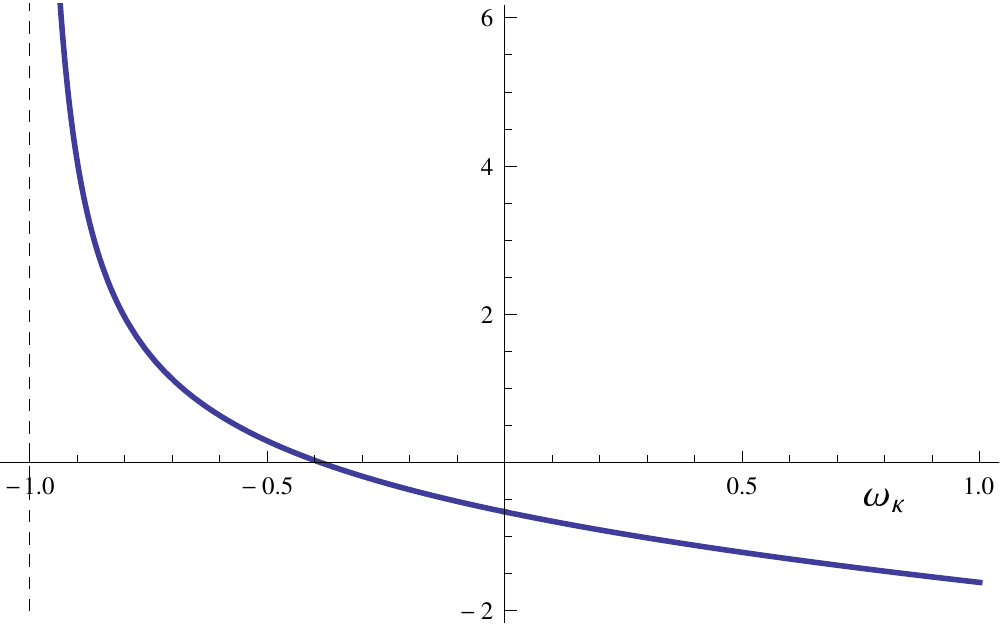} 
\caption{\label{righthandside}Right-hand-side of the equation (\ref{sol3FlargeNthetareg}) as function of 
$w_k$.}
\end{figure}
\begin{figure}[ht]
\centering\includegraphics[width=8cm]{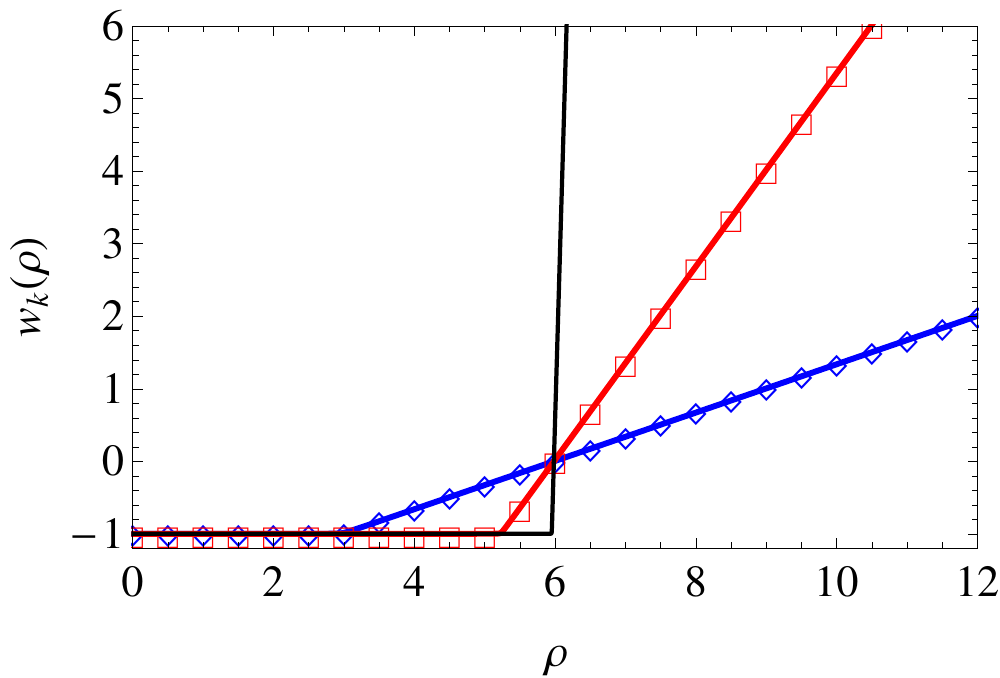} 
\caption{\label{solFlargeNthetareg} (color online) Solution of equation 
(\ref{sol3FlargeNthetareg}) for $w_k(\rho)$ as a function of $\rho$ for various
values of $k=10^{-2}\Lambda$ (blue diamonds), $k=10^{-3}\Lambda$ (red squares), $10^{-4}\Lambda$ (black plain line).}
\end{figure}
\begin{figure}[ht]
\centering\includegraphics[width=8cm]{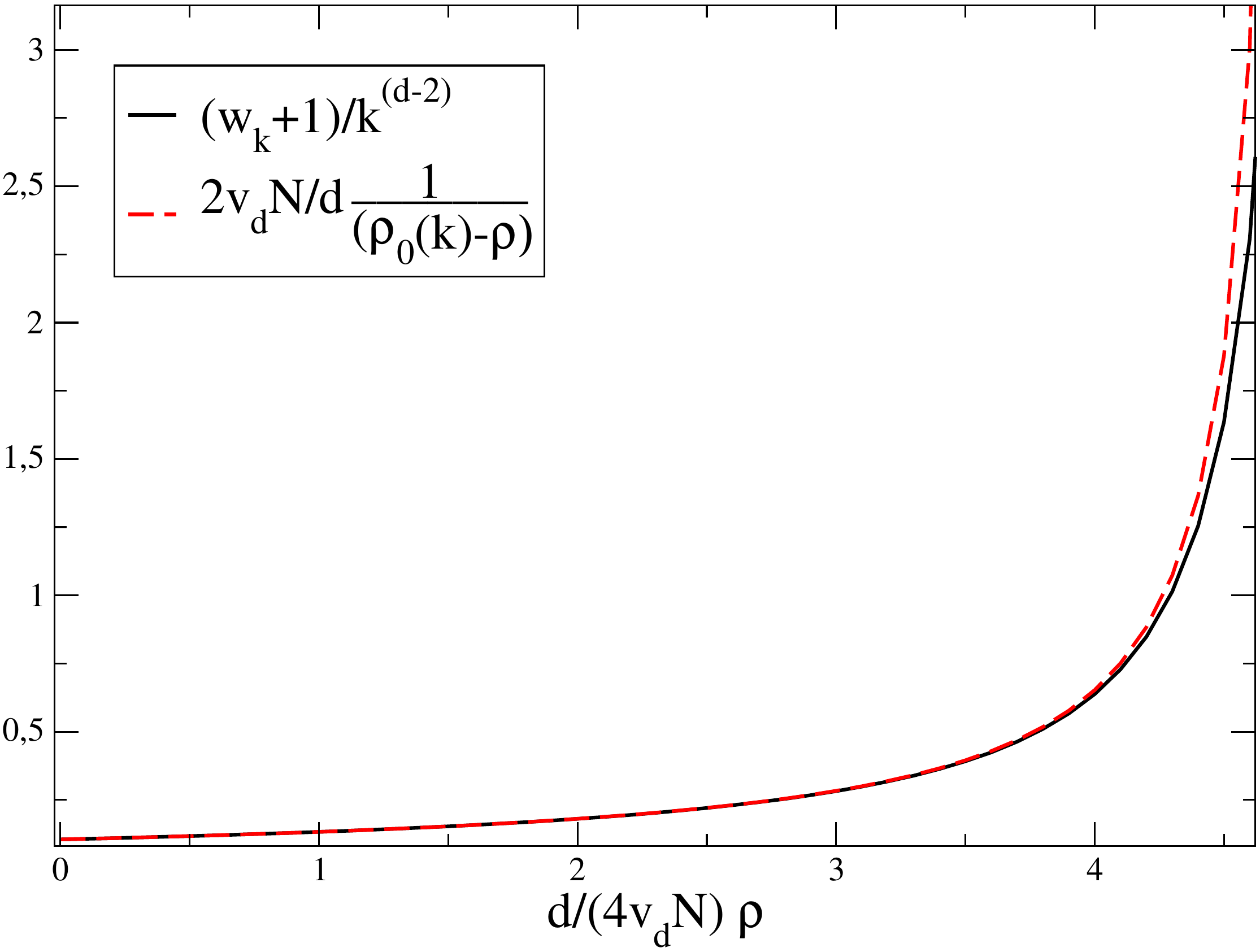} 
\caption{\label{dwkNinfLitim} (color online) Comparison of numerical (black plain line) and semi-analytical (red dashed)
solutions for $w_k+1$ at large $N$ as a function of $\rho$ for the 
$\theta$-regulator ($d=3$) for $k=0.002$.}
\end{figure}

Here we used as before the notation $w_k(\rho)=W_k(\rho)/k^2$. As for the flow equation (\ref{lpaeqthetareg}), right hand side of (\ref{sol3FlargeNthetareg})
when $w_k(\rho)\to -1$. This implies that $w_k(\rho)=-1$ plays the role of a 
barrier and the solution never goes reaches it. In Fig.\ref{righthandside}, the right-hand-side of
Eq.~(\ref{sol3FlargeNthetareg}) is plotted as a function of $w_k$, and in Fig. 
\ref{solFlargeNthetareg}
 the numerical solution of the implicit Eq.~(\ref{sol3FlargeNthetareg}) is shown 
for typical parameters in the
low temperature phase. One observes that the singularity is approached by the 
solution in the internal region of the potential. Moreover, in Fig.~\ref{dwkNinfLitim}, it can be
seen that the singularity is approached but is not crossed. This is very similar to the results obtained in \cite{Berges02} 
except that it was not known that the
$\theta-$regulator leads excatly to (\ref{sol3FlargeNthetareg}). In 
fact, the approach of the singularity can be discussed analytically.
First of all, the right hand side of
Eq.~(\ref{sol3FlargeNthetareg}) is a monotonous decreasing function. Accordingly 
it is not difficult to convince oneself that a unique solution exists for any $\rho$ and $k$. This means that the singularity 
is never reached. Second, if the singularity is not crossed
and a low temperature phase exists, there are values of $\rho$ with $w_k<0$ for 
all $k>0$. There are then only two possibilities. The first one, that
 corresponds either to $w_k\to 0^-$ or or to a negative constant larger than $-1$ in the internal region.
However, if this were true, the right hand 
side of Eq.~(\ref{sol3FlargeNthetareg}) would tend to a constant and the left hand side would tend
to infinity when $k\to 0$ giving a contradiction.
Correspondingly, the only remaining
possibility is that $w_k(\rho)$ tend to $-1$ in all the internal region of the 
effective potential.
Consequently, one can make an expansion of Eq.~(\ref{sol3FlargeNthetareg}) in $\delta w_k(\rho)=w_k(\rho)+1$.
At leading order one obtains:
\begin{equation}
\label{sollargeNIR}
 \delta w_k(\rho)= \frac{2 v_3 N}{3}\frac{k}{-3 \tilde r/u-\rho} 
\end{equation}
that, as observed before, leads to $\delta w_k(\rho)$ going to zero when $k\to 0$.

One can repeat this calculation for arbitrary integer dimension, but it is 
convenient to generalize it to an arbitrary $d$ by performing the expansion
on $\delta w_k(\rho)$ directly at the level of flow equations. This allows 
the generalization of this procedure to arbitrary values of $N$.
Before doing that, let us show the corresponding result for large values of $N$. If one 
expands at leading order on $\delta w_k(\rho)$ the flow equation
(\ref{lpaeqthetareg}) (taken at large $N$) one arrives at
\begin{equation}
 \Big( \frac 1 {\delta w_k(\rho)}\Big)'\sim- \frac{d k^{2-d}}{2 v_d N}
\end{equation}
whose solution is
\begin{equation}
\label{solthetareg}
 \delta w_k(\rho)= \frac{2 v_d N k^{d-2}}{d} \frac{1}{\hat \rho_0(k)-\rho}.
\end{equation}
Here $\hat \rho_0(k)$ is a integration constant that cannot be fixed without referring 
to the full (analytical or numerical) solution from
the microscopic scale $\Lambda$ to the infrared limit ($k\to 0$). In the 
particular large $N$ case, this constant can be fixed analytically
as in (\ref{sollargeNIR}). In the $k\to 0$ limit, moreover, it must be 
identified with $\rho(k=0)$ because, as discussed before, in the entire internal zone of the 
potential, the approximation just analyzed becomes correct when $k\to 0$.
From the solution of this equation we observe that the limit of validity 
of this approximation is precisely $\rho<\rho_0(k=0)$ when $k\to 0$.
Another consequence of this general solution is that there is no broken phase in 
LPA for $d\leq 2$ (as expected from the Mermin-Wagner theorem).
For $d\leq 2$, the ``correction'', does not tend to zero, and the associated 
solution does not exist. Following the previous discussion, the only
possibility, in absence of singularities is that at a given $k_0>0$, the minimum 
of the effective potential
reaches $\rho=0$ and remains there after.

The results from the numerical solution of (\ref{lpaeqthetareg}) (taken at large 
$N$) coincides with the previous results. One can solve
the equation with a standard finite differences explicit Euler procedure (with typical parameters $\Delta 
\rho=0.034$ and $\Delta t=-10^{-5}$ respectively). In 
Fig.\ref{dwkNinfLitim} the corresponding results are shown. Both solutions agree with good precision for
values of $\rho$ for which $w_k(\rho)<0$.
However, the singularity is approached and eventually the numerical code 
brings the potential to the wrong
side of the singularity and the flow blows up. This is a purely numerical problem. In fact, by improving the parameters
of the numerical code one can push the flow to smaller values of $k$. However, given that the singularity is approached rapidly
it becomes impossible to go to really small values of $k$ by simply taking smaller grids and larger volumes in $\rho$.
To solve this problem, we will present below an improved 
numerical algorithm that solves this difficulty.

\begin{figure}[ht]
\centering\includegraphics[width=8cm]{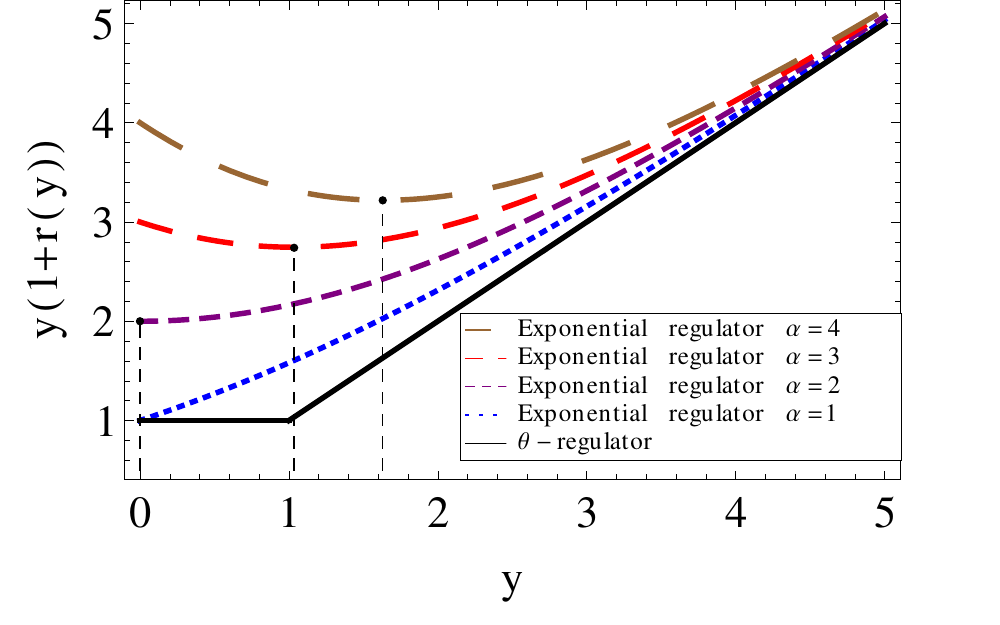} 
\caption{\label{regymin} $y(1+r(y))$ for different regulators. The vertical dashed lines show the minimun of $y(1+r(y))$, $y_0$, for each case.}
\end{figure}

The large-$N$ limit of the LPA equation (\ref{lpalargeN}) and its solution 
(\ref{sol3FlargeN}) has been partially analyzed previously for smooth regulators 
as the
exponential one \cite{Tetradis92,Berges02}. In fact, there are essentially two 
typical cases (see Fig.~\ref{regymin}): case (i) the inverse propagator $y(1+r(y))+W_k(\rho)/k^2$
has its minimum at a non zero value of $y$ (let us call it $y_0$), and case (ii) 
the inverse propagator has its minimum at $y=0$ with the
derivative of the inverse propagator with respect to $y$ being positive at 
$y=0$.
For the exponential regulator (\ref{expreg}) the case (i) corresponds to the 
case $\alpha>2$ (i.~e. for a strong enough regulator) and
the case (ii) corresponds to the case $\alpha<2$. There is a third possible case 
that corresponds to a minimum of the inverse propagator at $y=0$, the
derivative of which is zero. That is a very peculiar possibility that should be 
analyzed case-by-case.

\begin{figure}[ht]
\centering\includegraphics[width=8cm]{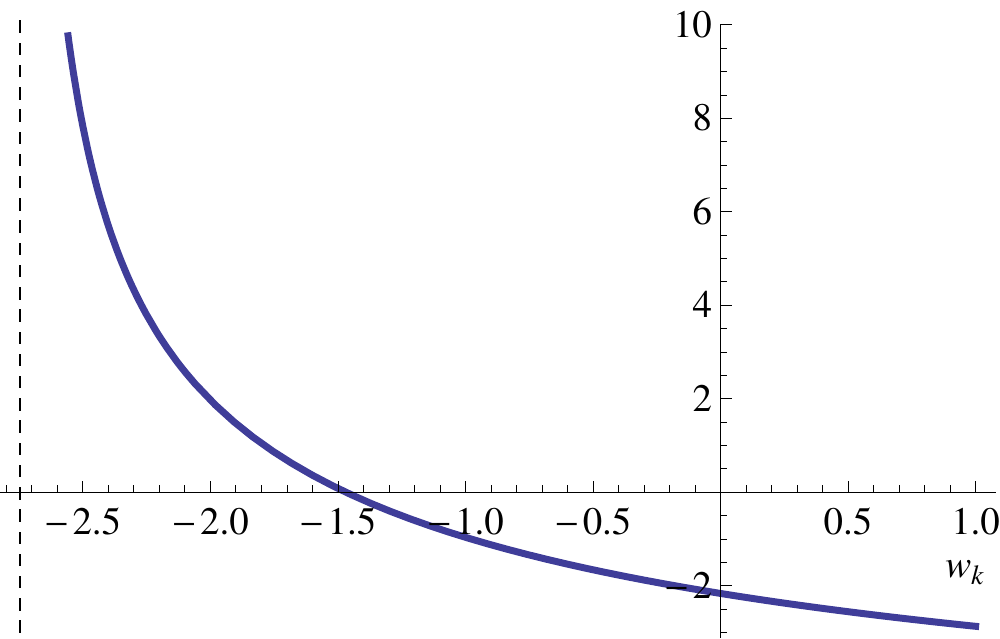} 
\caption{\label{largeNalpha3} Right hand side of equation (\ref{sol3FlargeN}) 
for the exponential regulator and $\alpha=3$ ($d=3$). The dotted line point out the position
of the singularity $-y_0(1+r(y_0))$.}
\end{figure}

Let us discuss first case (i). It has been shown that in this case, the 
behavior of the flow in the low temperature phase
is qualitatively similar to the one analyzed for the $\theta$-regulator 
\cite{Tetradis92,Berges02}: the singularity works as a barrier that is 
approached but never crossed
and accordingly the convexity of the effective potential is ensured by the LPA 
equation. The exponent characterizing the approach to the
singularity does not depend on the specific form of the regulator profile but is 
different to the particular case of the $\theta$-regulator.
The right-hand side of Eq.~(\ref{sol3FlargeN}) diverges
when $W_k(\rho)/k^2$ approaches $-y_0(1+r(y_0))$. It is not hard to convince 
oneself
that this singularity comes from the region of integration $y\approx y_0$. This 
can be seen in Fig~\ref{largeNalpha3} where the right hand side of
Eq.(\ref{sol3FlargeN}) is represented in the case $\alpha=3$.

Accordingly, one can obtain the equivalent of the integral when the
singularity is approached by substituting in the numerator of the integral $y$ 
by $y_0$ and by expanding the denominator at leading non trivial order
(at order $(y-y_0)^2$). Eq.~(\ref{sol3FlargeN}) near the singularity becomes:
\begin{align}
\label{sol3FlargeNsing}
\rho-\frac 3 u (W_k(\rho)-\tilde{r})&=-v_d N k^{d-2}\int_{-\infty}^\infty 
\frac{dy\,y_0^{d/2-1} }{\delta w_k+C (y-y_0)^2}\nonumber\\
&=-v_d N k^{d-2}\frac{\pi\,y_0^{d/2-1} }{\sqrt{C\,\delta w_k}}
\end{align}
where the notations $\delta w_k=y_0(1+r(y_0))+W_k(\rho)/k^2$ and $C=r'(y_0)+y_0 
r''(y_0)/2$ have been introduced. In this equation, the integration domain
has been enlarged from $-\infty$ because this integration domain is regular in 
the limit $\delta w_k\to 0$. It is important to observe that the $(\delta w_k)^{-1/2}$
behavior does not depend on the precise shape of the 
regulator as long as it has a regular behavior around $y_0$, the minimum at non 
zero
value of $y$ and as long as $C$, the second derivative of the inverse propagator 
at $y_0$ is non-zero. This second hypothesis is not fulfilled by
the $\theta$-regulator and this is why the right-hand-side of Eq.~(\ref{sol3FlargeN}) has a different 
behavior. In fact, when $n-1$ derivatives
of the inverse propagator with respect to $y$ are zero at $y_0$, the behavior of the right-hand-side of Eq.~(\ref{solFlargeN}) is 
as $(\delta w_k)^{-1+1/n}$, the $\theta$-regulator corresponding to
the limit $n\to \infty$. 

Eq.~(\ref{sol3FlargeNsing}) can now be inverted by observing that, for the same reasons invoked 
for the $\theta$-regulator that, the singularity is approached but
never reached. Accordingly when $k\to 0$, one can expand the Eq.~(\ref{sol3FlargeNsing}) on $\delta w_k$. At leading order, one obtains:
\begin{equation}
\label{solsmoothreg}
 \delta w_k(\rho)= \frac{1}{C}\Bigg(\frac{v_d N k^{d-2}\pi 
y_0^{d/2-1}}{\rho_0-\rho}\Bigg)^2 
\end{equation}
with $\rho_0=-3\tilde r/u$.
As done for the $\theta$-regulator, one can also obtain a similar expression by 
integrating directly the flow equation (\ref{lpalargeN}). One obtains
the same expression, except that $\rho_0$ is replaced by an arbitrary function 
of $k$, $\hat \rho_0(k)$ that comes as an integration constant (independent
of $\rho$). As 
before, in the limit $k\to 0$, $\hat \rho_0(k)$ can be interpreted as the position
of the minimum of the effective potential $\rho_0(k=0)$.

\begin{figure}[ht]
\centering\includegraphics[width=8cm]{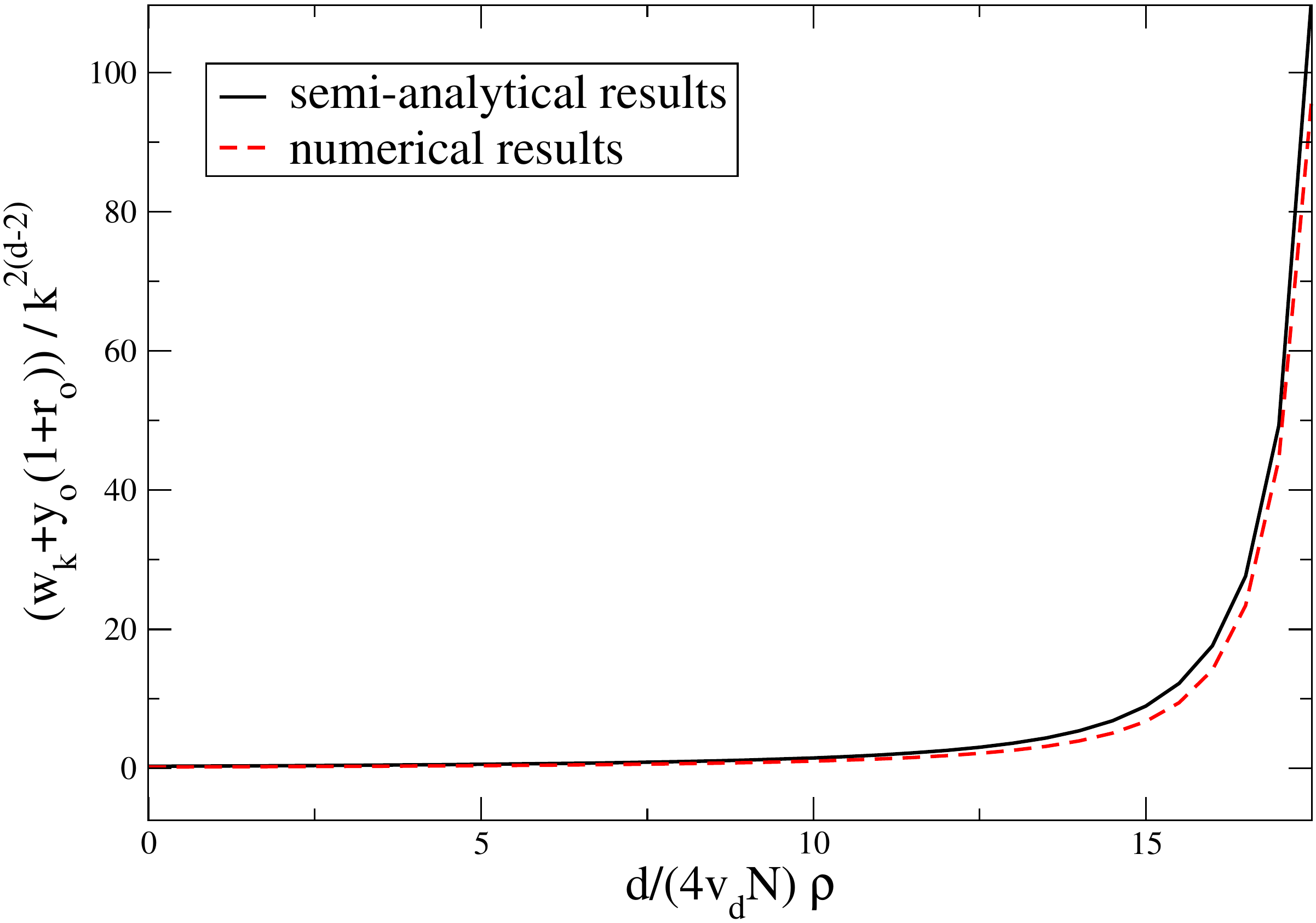} 
\caption{\label{dwkNinfExp} (color online) Comparison of numerical (black plain line) and semi-analytical (red dashed)
solutions at large $N$ for $w_k+y_0(1+r(y_0))$ as a function of $\rho$ for the 
exponential regulator
($d=3$). Curves for $k=0.08$.}
\end{figure}

We display in Fig.~\ref{dwkNinfExp} a numerical solution of the LPA equation in the large $N$ 
limit (\ref{lpalargeN}) that has been, as before,
using finite differences and explicit Euler method with typical parameters $\Delta 
\rho=0.034$ and $\Delta t=-10^{-5}$ respectively. 
It must be stressed that to observe numerically the proper behavior of the solution 
in the internal region of the potential is much more numerically demanding than, for example, to study of the critical 
behavior of these models (in that case one can typically obtain stable results with
$\Delta \rho/k^{d-2}=0.1$ and $\Delta t=-2.10^{-3}$).
The only difference with the $\theta-$regulator is that the integrals over 
momenta must be done
numerically. We employ for this purpose Simpson's rule with a regular grid in 
momenta with 80 steps of a dimensionless momentum step of 0.1.
The solution agrees
with the analytical behavior just presented. However, as with the 
$\theta$-regulator, at a certain value of $k$, the singularity is crossed due to
numerical lack of precision and consequently the flow collapses. As with the $\theta$-regulator,
we present below a more elaborated method in order to avoid
such collapse.

\begin{figure}[ht]
\centering\includegraphics[width=8cm]{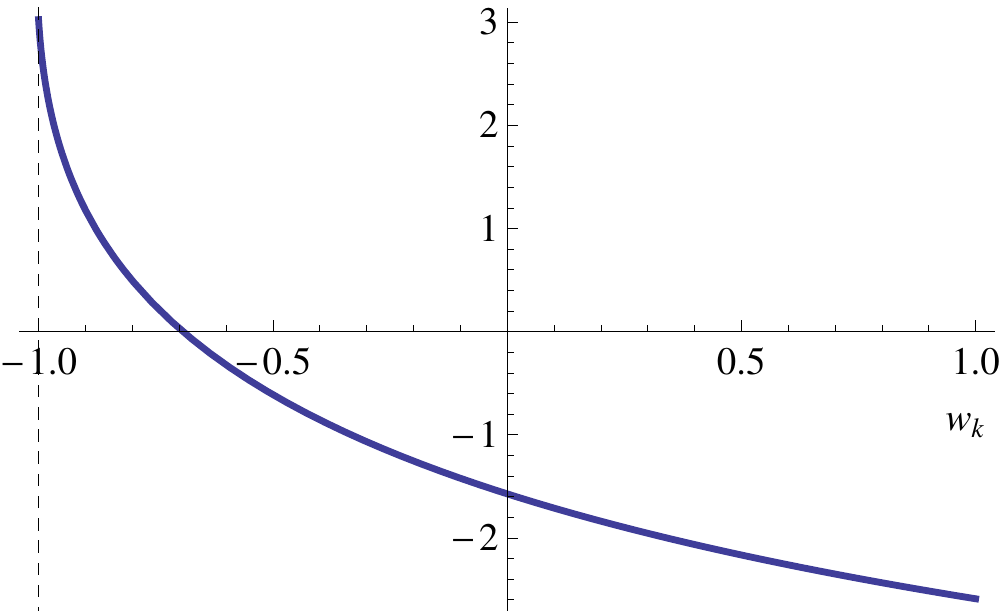} 
\caption{\label{largeNalpha1} Right hand side of equation (\ref{sol3FlargeN}) 
for the exponential regulator and $\alpha=1$ ($d=3$). The dotted line point out the position
of the singularity $-\tilde R=-\lim_{y\to 0} y (1+r(y))$}
\end{figure}

\begin{figure}[ht]
\centering\includegraphics[width=8cm]{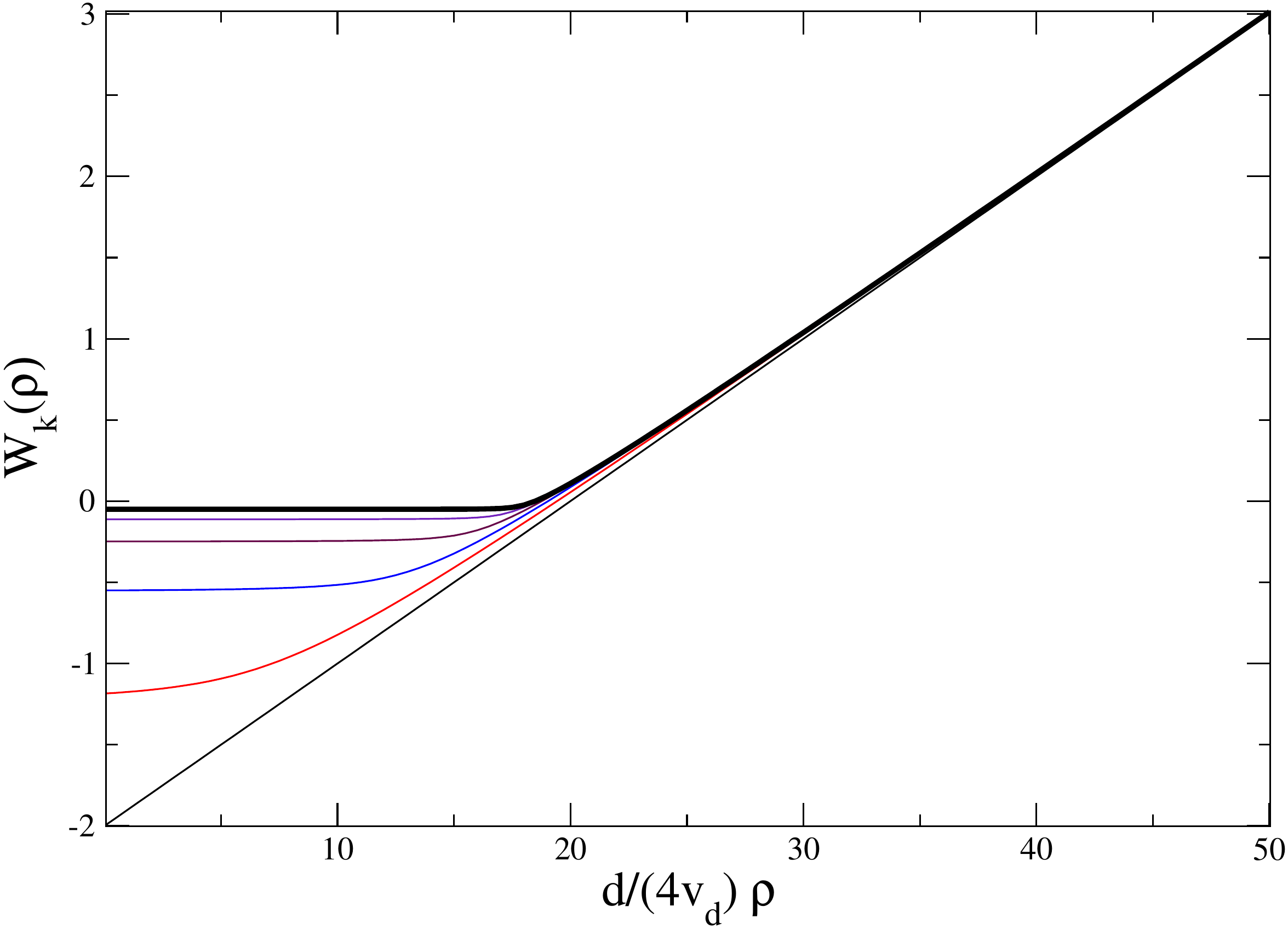} 
\caption{\label{WkNinfExp} Derivative of the potential at large $N$ and $d=3$ as 
a function of $\rho$ for various values of $k$ (lower curves are for larger 
values of $k$).}
\end{figure}

Let us now consider the case (ii) (corresponding to $\alpha<2$ in the particular 
case of the exponential regulator).
In this case, the singularity does not work any more as a barrier and the 
integral remains bounded when the singularity is approached. The singularity
in that case shows up first at $y=0$, and corresponds to the point where 
$w_k(\rho)$ approaches $-\tilde R\equiv-\lim_{y\to 0} y(1+r(y))$. The integral
is not differentiable at $w_k(\rho)=-\tilde R$ but remains continuous at this 
point. In the particular case of the expontial regulator (\ref{expreg}), $\tilde R=\alpha$.
As mentioned before, the flow blows up at a finite scale $k_0$ because $U_k$ hits the singularity at$k=k_0$.
This singularity that occurs at finite $k$  is also observed when numerically integrating the flow equation.
However, this is not very conclusive because when $\alpha>2$, the singularity that should not be reached
in principle is actually reached because of numerical inaccuracies.
However, in order to be fully 
convinced that the singularity
is hit, one can exploit the implicit large $N$ solution (\ref{sol3FlargeN}) and 
observe that there is a solution for $w_k=-\tilde R$ when $\rho$
and $k$ are small enough. In order to see that, one can observe that the 
right-hand-side of equation (\ref{sol3FlargeN}) is
bounded from above as a function of $w_k$. As an example, in 
Fig.~\ref{largeNalpha1}, the case $\alpha=1$ is represented. One sees that the
right-hand-side presents a singularity but that it is finite, not diverging.
As a consquence, nothing forbids the large $N$ implicit solution (\ref{sol3FlargeN})
to reach the singularity at $k>0$.
Having discussed the standard numerical solution of the equation, the implicit 
analytical solution and the explicit analytical solution near the
singularity, we present now an improved numerical solution that exploits 
the obtained analytical behavior in the two cases discussed above
where the singularity is avoided: the $\theta$-regulator and the smooth 
regulator in case (i). The idea is simple and has been already suggested (but
not implemented) in \cite{Berges02}. One can employ a standard numerical 
procedure at typical points in a grid, but in a region where the solution is 
close enough to the singularity (and where the analytical solutions (\ref{solthetareg}) or (\ref{solsmoothreg}) are therefore justified),
 one  replaces the result of the flow equation by the analytical expressions
(\ref{solthetareg}) or (\ref{solsmoothreg}),
depending on the chosen regulator. For a given smooth regulator, the constant 
$y_0$ can be calculated (by looking at the minimum of $y (1+r(y))$).
For values of 
$\rho$ at which $w_k$ is above a chosen threshold, one implement
a standard numerical solution (finite differences plus explicit Euler). The 
value of the integration constant $\hat \rho_0(k)$ is taken in order to require the continuity between
the analytical solution below the threshold and the purely numerical one above 
it.
It must be stressed that this algorithm requires the knowledge of the solution 
in the full internal region and not only around $\rho=0$ as was obtained in
\cite{Tetradis92,Tetradis95}. For actual numerical implementations with  
the $\theta$-regulator we took the value for the threshold at $w_k=-0.98$.
In the case of the the exponential regulator, we chose $\alpha=3$ (for which 
$y_0\simeq1.035658$) and we chose the threshold value $w_k=-2.74$. 
This numerical procedure is completely stable. The flow can be continued down to $k/\Lambda\sim 
1.5\times 10^{-8}$ without encountering any difficulty. From the result of $w_k$ 
one
can reconstruct the dimensionful potential which, as expected, is convex. The 
corresponding result is shown in Fig.\ref{WkNinfExp}.

The large $N$ limit is not particularly exciting because the flow equation can 
be essentially solved analytically and because physically interesting
models correspond to lower values of $N$. We merely use it to test various ideas about the approach to convexity.
In the following we exploit such ideas to more realistic 
values of $N$, beginning with the $N=1$ and finally generalizing to
other values of $N$.

\subsection{Finite $N$}

We analyze now the the finite $N$ case, where no (even 
implicit) analytical solution is known. As done for large $N$ we consider
the corresponding equations (\ref{lpaeqthetareg}) and (\ref{lpaW}) both analytically and numerically, first with the 
$\theta$-regulator and then for a generic smooth regulator
(the corresponding numerical implementation is performed for the exponential 
one).

Consider first the LPA flow equation with the $\theta$-regulator for $N=1$:
\begin{equation}
\label{lpaeqthetaregNeq1}
\partial_tw_k=-2w_k-\frac{4 v_d}{d}k^{d-2}\frac{3w_k'+2\rho w_k''}{(1+w_k+2\rho 
w_k')^2}
\end{equation}
Again, let us admit (as clearly seen in the numerical solution of the equation 
in $d=3$) that the LPA equation displays a low temperature phase in
which there is a minimum of the potential $\rho_0(k)$ with 
$w_k(\rho_0(k))=W_k(\rho_0(k))/k^2=0$ for any $k>0$ and with $\rho_0(k=0)>0$. Repeating
a similar analysis to the one performed for large $N$, one concludes that the
flow approaches the singularity where
$1+w_k(\rho)+2\rho w'_k(\rho)=0$. In that case, there are, again two possibilities: 
either the singularity is never reached or it is crossed.

It is difficult to
give a general analytical proof for finite $N$ that the singularity is not 
crossed at any finite value of $k$, but the numerical solution of the equations
gives clear indications in this direction for the $\theta-$regulator. Under this hypothesis,
one concludes that when $k$ is small enough the flow approaches a regime when 
$1+w_k(\rho)+2\rho w'_k(\rho)$
is small in all the internal region of the potential but positive.
Moreover, in absence of singularities for $k>0$, $w_k(\rho)$ is a regular 
function of $\rho$. The solution of the equation
$1+w_k(\rho)+2\rho w'_k(\rho)\approx 0$ is
\begin{equation}
w_k(\rho)\approx -1 + 
A_k/\sqrt{\rho},
\end{equation}
where $A_k$ is an arbitrary function of $k$. However, the 
solution
being regular for any $\rho$ and, in particular for $\rho=0$, one concludes that 
$1+w_k(\rho)+2\rho w'_k(\rho)\approx 0$ is equivalent to 
$1+w_k(\rho)\approx 0$ in the entire internal region of the potential. This is 
the same behavior as for large $N$ but for a slightly subtler reason.
Moreover, as when $N$ is large, one can analyze the approach to this regime by 
expanding Eq.~(\ref{lpaeqthetaregNeq1}) in $\delta w_k(\rho)=w_k(\rho)+1$:
\begin{equation}
\label{expansionwk}
\frac{4 v_d}{d}k^{d-2}\frac{3 \delta w_k'+2\rho \delta w_k''}{(\delta w_k+2\rho 
\delta w_k')^2}=2+\mathcal{O}(\delta w_k)
\end{equation}
Neglecting the term $\mathcal{O}(\delta w_k)$ in the right hand side, one can solve the Eq.~(\ref{expansionwk}). The 
solutions that are regular at $\rho \sim 0$ are of
the form
\begin{equation}
 \label{analytsolN=1}
 \delta w_k(\rho)=\frac{v_d}{d}\frac{k^{d-2}}{\sqrt{\rho_0(k)\rho}} 
\log\Bigg(\frac{\sqrt{\rho_0(k)}+\sqrt{\rho}}{\sqrt{\rho_0(k)}-\sqrt{\rho}}
\Bigg)
\end{equation}
where $\rho_0(k)$ is an arbitrary function depending on the initial conditions 
of the flow. Given that this solution is only
valid for small values of $k$, a possible $k$ 
dependence of $\rho_0$ can be neglected. It is in the $k\to 0$ limit the position of the minimum of the 
effective potential.

\begin{figure}[ht]
\centering\includegraphics[width=8cm]{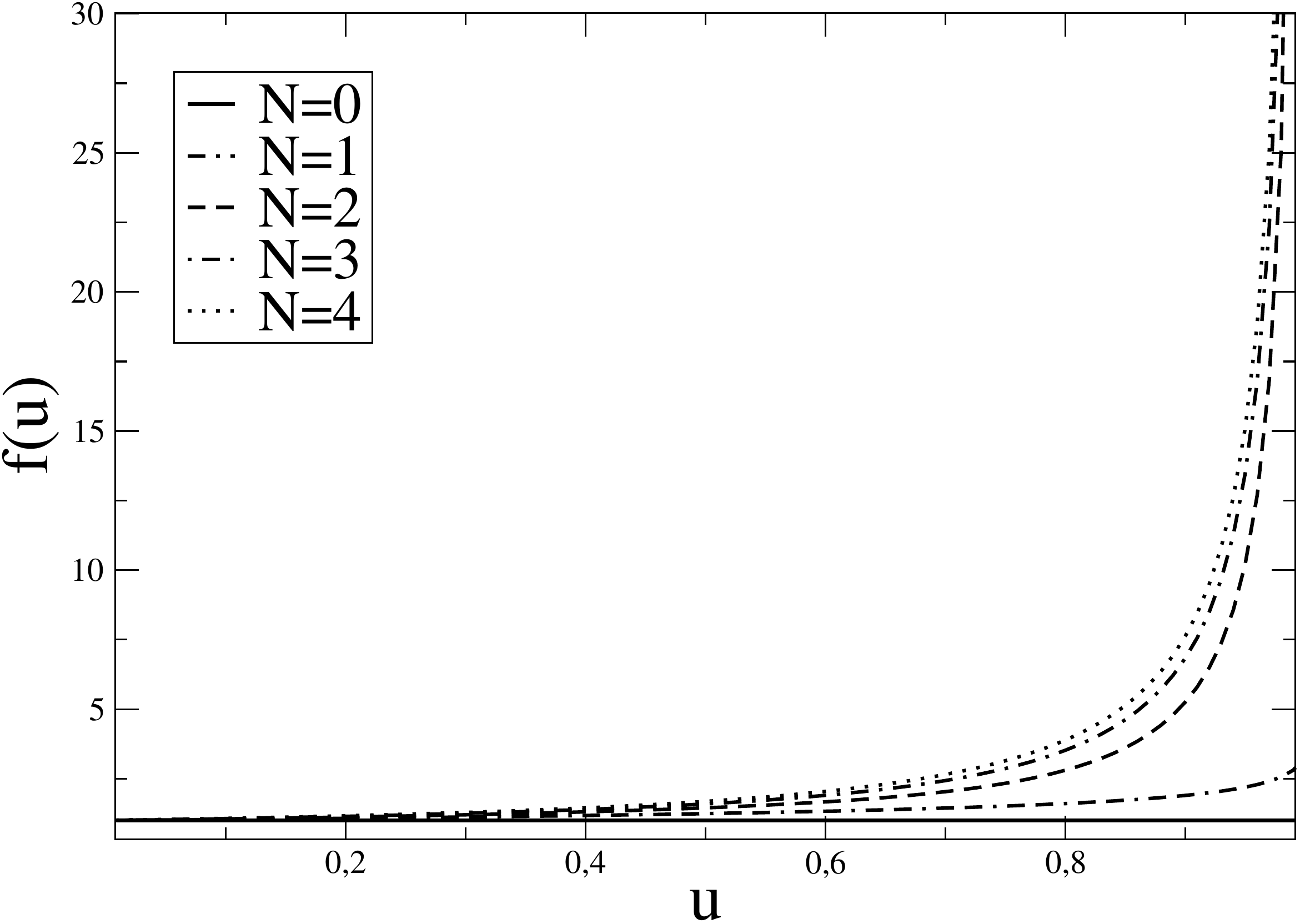} 
\caption{\label{fdeuN} $f(u)$ as a function of $u$ for various values of $N$.}
\end{figure}

For generic values of $N$, the LPA equation with the $\theta$-regulator is:
 \begin{equation}
\label{lpaeqthetaregdimensionless}
\partial_tw_k=-2w_k-\frac{4 v_d}{d}k^{d-2}\Bigg(\frac{3w_k'+2\rho 
w_k''}{(1+w_k+2\rho w_k')^2}+\frac{(N-1)w_k'}{(1+w_k)^2}\Bigg)
\end{equation}
The two convexity conditions to be fulfilled are those of large $N$ and of 
$N=1$. As explained before, and admitting that
both singularities are not crossed at finite $k$, both of them imply that when $k$ is small, 
$\delta w_k(\rho)=w_k(\rho)+1 \ll 1$.
As in previous cases, one can expand the equation 
(\ref{lpaeqthetaregdimensionless}) in $\delta w_k$ yielding the differential 
equation
 \begin{equation}
 \label{eqdwgenericN}
\rho_0(k)-\rho=\frac{2 v_d}{d}k^{d-2}\Bigg(\frac{1}{\delta w_k+2\rho \delta 
w_k'}+\frac{N-1}{\delta w_k}\Bigg)
\end{equation}
where $\rho_0(k)$ is an arbitrary function depending on initial conditions. As 
before, $\rho_0(k)$ can be interpreted when $k\to 0$ as the position
of the minimum for the potential, and one can neglect its $k$ dependence.
Eq.~(\ref{eqdwgenericN}) cannot be solved analytically except for the 
previously considered cases ($N=1$ and large $N$
\footnote{In fact, at $N= 0$ case can be handled analytically also. In that 
case, a $\delta w_k$ independent of $\rho$ is solution of (\ref{eqdwgenericN}).
If one ask for the continuity of the solution when $N\to 0$, the solution is 
(see (\ref{regularitycond})) completely fixed.}). Being a differential
equation one could expect that for any $\rho_0$ it has an infinite number of 
solutions corresponding to different choices of $\delta w_k(\rho=0)$.
However, as before, one must require that it is well-behaved in all the domain of 
validity of the approximation, and in particular for $\rho=0$.
This 
fixes the value of $\delta w_k(\rho=0)$ in terms of $\rho_0$:
 \begin{equation}
 \label{regularitycond}
\delta w_k(\rho=0)=\frac{2 v_d}{d}k^{d-2}\frac{N}{\rho_0(k)}
\end{equation}
yielding a single regular solution in the domain of validity of the equation. The 
Eq.~(\ref{eqdwgenericN}) can be solved numerically easily. It is convenient to define
\begin{align}
 u&= \frac{\rho}{\rho_0(k)}\nonumber\\
 f(u)&=\frac{\delta w_k(\rho)}{\delta w_k(\rho=0)}
\end{align}
that yields the following equation for $f(u)$:
\begin{equation}
\label{quasi-analyticalbis}
1-u=\frac{1}{N}\Bigg(\frac{1}{f(u)+2 u f'(u)}+\frac{N-1}{f(u)}\Bigg)
\end{equation}
with the initial condition $f(u=0)=1$. The expression of $\delta w_k(\rho)$ can be reconstructed
from that of $f(u)$ obtained at a given $N$ and for an arbitrary $\rho_0(k)$:
\begin{equation}
 \label{quasi-analytical}
 \delta w_k(\rho)=\frac{2 v_d}{d}k^{d-2}\frac{N}{\rho_0(k)}f(\rho/\rho_0(k))
\end{equation}
The form of $f(u)$ for various values of $N$ obtained by numerically solving the 
Eq.~(\ref{quasi-analyticalbis}) are shown in Fig.\ref{fdeuN}. 
It must be stressed that the {\it correction} to $\delta w_k \approx 0$ 
differs fromt its large $N$ limit. However, for any $N$ one 
generically
approaches the regime where  $\delta w_k \approx 0$ but the corresponding 
function $f(u)$ depends on $N$ for generic values of $\rho$ in the internal region.
This is in contrast to the $\rho\to 0$ limit where it has been shown 
\cite{Tetradis92,Tetradis95} that the large $N$ form is self-consistent for any 
$N$.

\begin{figure}[ht]
\begin{center}$
\begin{array}{cc}
\includegraphics[width=8cm]{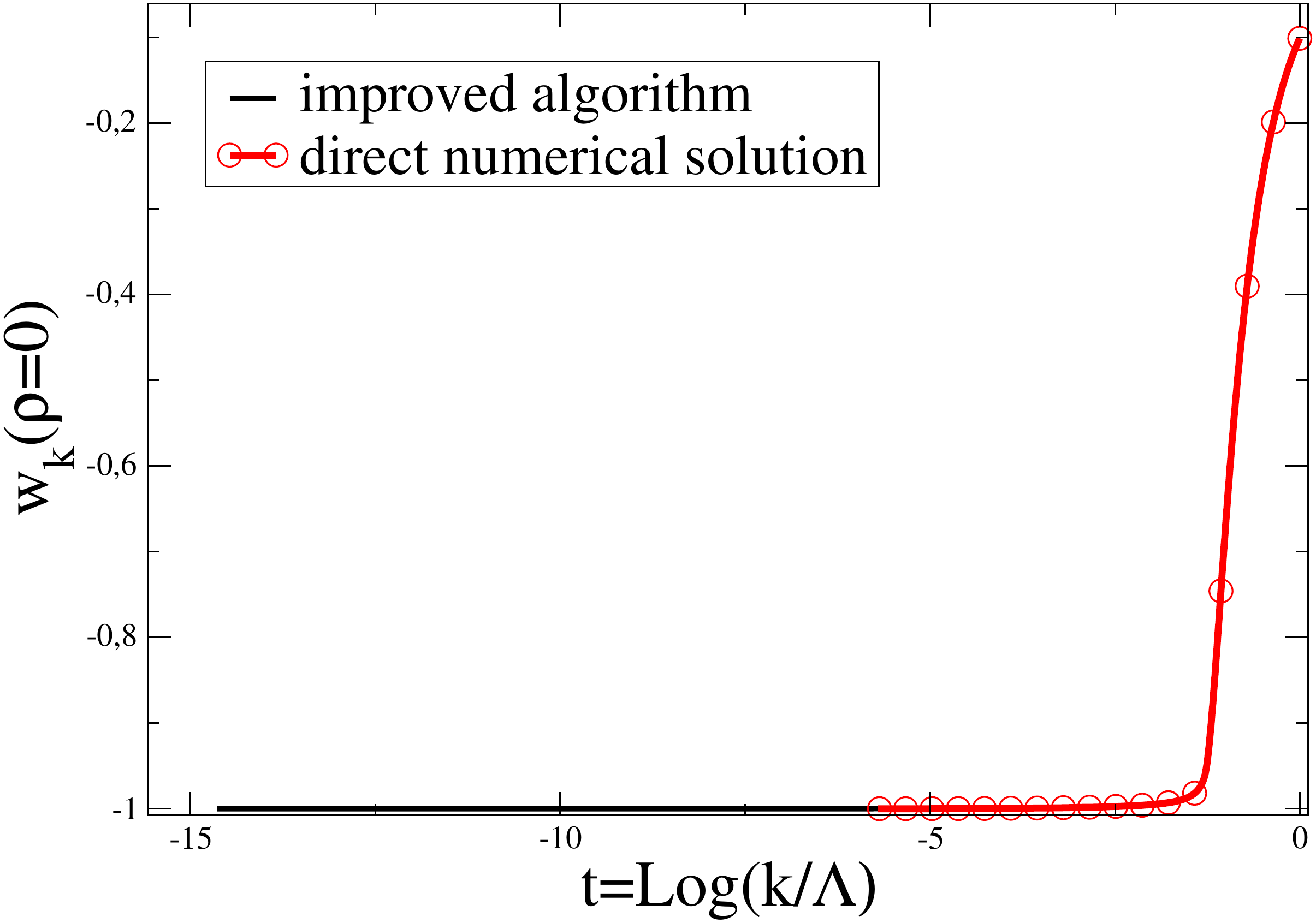}
\end{array}$
\end{center}
\caption{\label{FRAN4LPALitim} (color online) $w_k(\rho=0)$ as a function of 
$t=\log(k/\Lambda)$ for $N=4$ and $d=3$. Comparison between the direct
numerical solution (red circles) and the improved algorithm (black plain line) in
both cases with the $\theta-$regulator.}
\end{figure}

We also solve numerically the LPA equation with the $\theta$-regulator for 
various values of $N$ with the same procedure presented before for large $N$.
First, we solve it directly using finite-differences for the derivatives and an 
explicit Euler algorithm for the evolution in $t$. Typically the parameters
used are $\Delta\tilde\rho=0.1$ and $\Delta t=-10^{-5}$. As can be seen in 
Fig.~\ref{FRAN4LPALitim}, in all cases (including $N=0$), a low temperature 
phase is found for $d>2$ and the numerical solution indicates
that the function $w_k(\rho)$ approaches (without crossing) $-1$ in all the 
internal region of the potential. However, like for large $N$, when the solution
is too close to the singularity it may happen that discretization errors leads to an 
artificial crossing of the singularity.

It is interesting to note that the numerical implementation of the LPA equation 
for $N=1$ turns out to be much more difficult than for $N>1$.
The reason for this, at first sight surprising, result is two-fold. First, as explained before, for $N>1$ the convexity condition
$w_k(\rho)>-1$ is imposed directly by the term of the LPA Eq.~(\ref{lpaW})
proportional to $N-1$. The other term of the right-hand-side of the
equation (the only present when $N=1$), imposes the weaker constraint 
$w_k(\rho)+2\rho w_k'(\rho)>-1$ that eventually leads to the same 
consequence
($w_k(\rho)>-1$) but in a much more indirect way (see above). Numerically this effect seems to be harder
to control. The second reason is that, for $N=1$, the dimensionful physical effective potential $U_{k=0}$ has a 
discontinuity in its second derivative at the minimum of the potential. This is 
simply
related to the fact that for $N=1$ the susceptibility is finite both in the high 
and low temperature phase (only diverging
asymptotically when the critical temperature is approached). On the contrary, 
the second derivative of the physical
effective potential is continuous for $N>1$, even at the minimum. This expresses
the fact that the susceptibility of the $O(N)$
models with continuous symmetries ($N>1$) is infinite for any temperature below 
the critical one because of Goldstone modes.
In practice, the effective potential for $k>0$ remains much more regular around 
the minimum for $N>1$ diminishing the sources of instabilities.

In order to improve the stability of the numerical solution, we employed the 
same procedure presented above for large $N$: we fixed a threshold $w^{\text{threshold}}=-0.95$,
solved numerically the flow equation (\ref{lpaW}) when $w_k(\rho)$ is above $w^{\text{threshold}}$ and imposed the 
quasi-analytical form given by Eq.~(\ref{quasi-analytical}) for $w<w^{\text{threshold}}$. The implementation of 
this procedure proves to be essentially stable at arbitrary values of $t$
for all $N>1$. A typical example of a such solution is shown 
Fig.\ref{FRAN4LPALitim}. As before, the numerical solution of Eq.(\ref{lpaW}) for $N=1$ is much 
more demanding.
In fact, the procedure explained above does not work for $N=1$ as efficiently as in the $N>1$ case,
although (\ref{expansionwk}) is again a good approximation in all the internal region of the potential.
It turns out that the mismatch between the second derivatives of the potential at the matching
point corresponding to $w^{\text{threshold}}$ is large enough to generate numerical instabilities.

\begin{figure}[ht]
\centering\includegraphics[width=8cm]{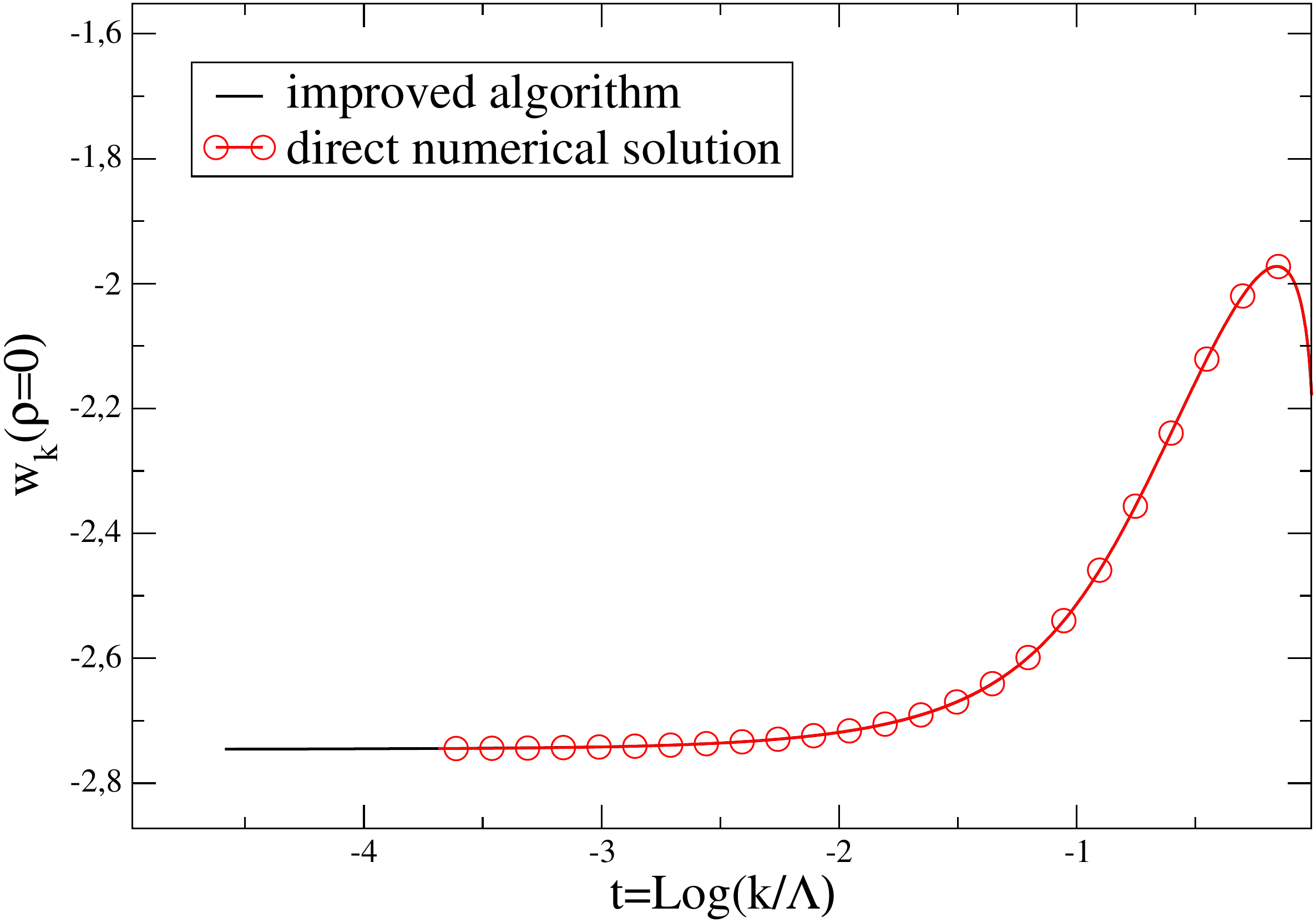} 
\caption{\label{WkNExp} (color online) Comparison between the direct
numerical solution (red circles) and the improved algorithm (black plain line).
$N=4$ and $d=3$ using the exponential regulator
with $\alpha=3$.}
\end{figure}

We analyzed the equation also for typical smooth regulators. Again, there are two different cases, depending on the position
of the minimum of $y(1+r(y))$. As for large $N$, when the minimum takes place at 
a $y=y_0>0$, integrals in the right-hand-side of the LPA equation
can be approximated as in (\ref{sol3FlargeNsing}). Accordingly the singularities 
play the role of a barrier that cannot be crossed and one arrives at
a scenario very similar to the one of the $\theta$-regulator: the singularity is 
approached but never crossed. 
In this case the behavior of the analytical solution when $k\to 0$ 
is
\begin{equation}
\delta w_k\sim k^{2(d-2)}f(\rho).
\end{equation}
For for $N=1$ the function $f(\rho)$ can be found analytically:
\[f(\rho)=\frac{1}{C^2}\Bigg(\frac{1}{\rho_0^{3/2}\sqrt{\rho}}\log\Big(\frac{
\sqrt{\rho}+\sqrt{\rho_0}}{\sqrt{\rho_0}-\sqrt{\rho}}\Big)+\frac{2}{
\rho_0(\rho_0-\rho)}\Bigg)\]
and for $N>1$ it can be obtained by solving the differential 
equation:
\[f+2\rho f'=\frac{1}{\Big(\frac C 2 
(\rho_0-\rho)-\frac{N-1}{\sqrt{f}}\Big)^2}.\]
In both cases the constant $C$ is related to the minimum as
$$C=\frac{y_0(1+r(y_0))}{v_d y_0^{d/2} r(y_0) 
\pi}\sqrt{r'(y_0)+y_0r''(y_0)/2}.$$
Following the same procedure we can go further in the solution of the flow 
equation as it is shown in the Fig.\ref{WkNExp}.
Having discussed the treatment of the LPA equation in the various cases and showing
a new numerical algorithm that is much more stable than the standard one for all
$N>1$, we consider now the next-to-leading order of the derivative expansion.

\section{Derivative Expansion at order $\mathcal{O}(\partial^2)$}

In this section, we generalize the previous numerical studies on the approach to a 
convex free-energy at the LPA level to second
order in the derivative expansion. This approximation corresponds to an 
expansion to second order of the NPRG equation in powers of the external 
momenta, see Eq.~(\ref{ansatzD2}). We derived the corresponding NPRG equations and we verified the equivalence with 
\cite{VonGersdorff:2000kp}.
We limit ourselves to a direct numerical analysis leaving for the future an 
analytical
study analogous to the one performed for the LPA. This analysis should again improve the numerical integration of the flow equations
but the number of cases to be studied brings it clearly beyond 
the scope of the present article.
One aspect that cannot be addressed at the LPA level is the broken phase for $N=1$ in $d=2$ since the running
of the field renormalization factor is neglected which artificially destroys 
the broken phase. On the contrary, at order $O(\partial^2)$ of the DE, a phase transition is found at finite
temperature for $N=1$ in $d=2$ \cite{VonGersdorff:2000kp,Jakubczyk:2014isa}. For $N>2$, no phase transition is found with this approximation in agreement
with the Mermin-Wagner theorem and for $N=2$, the Kosterlitz-Thouless phase transition is correctly
described \cite{Kosterlitz:1973xp}.\footnote{The cases with $N<1$ in any dimension and, in particular,
the physically interesting case $N=0$ require an independent analysis that goes
beyond the present article. In that case, the sign of the term in the potential
equation proportional to $N-1$ changes and, consequently, a different analysis 
is required.}

An important difference between the order $O(\partial^2)$ of the DE and the LPA
is that it is convenient to
introduce a pre-factor in the regulator function $R_k(q)$ that evolves with $k$ 
(as usually done in the study of the critical regime).
This pre-factor has many purposes in the critical regime.\footnote{For example, 
this pre-factor makes the fixed point condition of NPRG equations identical
to the Ward Identity of scale transformations in presence of an infrared 
regulator, see \cite{Delamotte:2015aaa}.} In the present case let us consider 
the
inverse propagator, (for $N=1$ for example):
\begin{equation}
G_k^{-1}(q)= q^2 Z_k(\rho)+W_k(\rho)+2\rho W'_k(\rho)+R_k(q).
\end{equation}
For $R_k(q)$ to regulate efficiently and for all values of $k$ the small
wave-number modes, it is necessary that it is at least of of the same order as
$q^2 Z_k(\rho)$ up to $q\sim k$. As usual, we use regulators of the form
\begin{equation}
 R_k(q)= Z_k \hat R_k(q)
\end{equation}
where $\hat R_k(q)$ are the regulator profiles used at the LPA level, see Eqs.~(\ref{thetareg},\ref{expreg}),
and $Z_k$ is fixed as $Z_k(\rho)$ at a particular value of $\rho$. The difficulty is that $Z_k(\rho)$ depends strongly on 
$\rho$ and, not surprisingly, the behavior of this function for $\rho$ larger or
smaller than the minimum of the potential is very different in the low 
temperature phase when $k\to 0$. We analyze two possible choices: $\rho$ 
larger
or smaller than $\rho_0$, the minimum of the potential when $k\to 0$. As we will 
see, the appropriate choice for this point depends on the value of
$N$. On one hand, when $N>1$, we observe that the flow is more stable if $Z_k$ 
is taken as the
value of $Z_k(\rho)$ for a $\rho>\rho_0$, in some cases in a very significant 
way. For this reason, for those values of $N$, all results presented
below correspond to this choice
of $\rho$ (more precisely, $\rho=2 \rho_0$). On 
the other hand, for $N=1$, one must fix the value of $Z_k$ for a $\rho$
in the ``internal'' part of the potential, as 
explained below. If this is not done, the flow of the potential hits the singularity as 
with the LPA for a regulator not strong enough.
As for the choice of the regulator profile, the main advantage of the $\theta$-regulator (\ref{thetareg})
is lost at the second order of the DE because the integrals cannot be performed any 
more analytically.
We therefore use the exponential regulator (\ref{expreg}) in what follows and we choose a prefactor
$\alpha$ larger than 2 to avoid singularities in the flow, see section \ref{secLPA}.

The large $N$ case is not particularly useful for the second order of the 
derivative expansion, because in that limit, the
LPA equation for the potential becomes exact. We consider then, first, the single 
scalar case, generalizing those results to
the $O(N)$ case after. An analysis of such theories has been done a few years 
ago at the second order of the derivative expansion
in \cite{Zappala12} but for $d=4$.
Here we consider dimensions $d<4$ that are generically
much richer for scalar theories. Moreover, an interesting but very weak logarithmic 
divergence of
the function $Z(\rho)+\rho Y(\rho)$ were observed in \cite{Zappala12}. In $d<4$,
we observe clearer and stronger effects because the corresponding 
divergences is power-law, as will be discussed below.

\subsection{Single scalar case}

\begin{figure}[ht]
\centering\includegraphics[width=8cm]{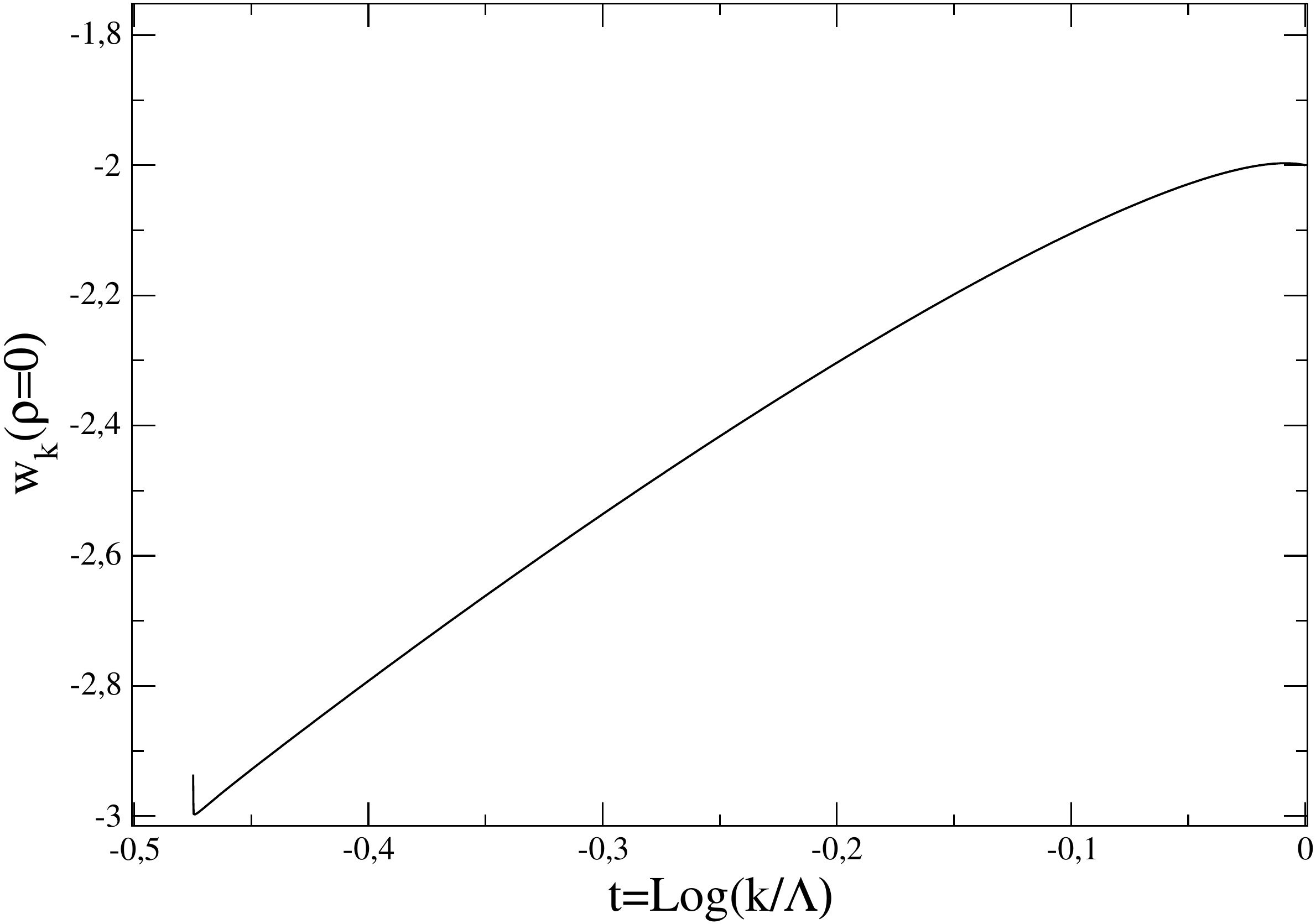} 
\caption{\label{wkD2N1normafuera} $w_k(\rho=0)$ as a function of 
$t=\log(k/\Lambda)$ for $N=1$ and $d=3$ using the exponential regulator (\ref{expreg})
with $\alpha=3$. Normalization fixed at $\rho=2\rho_0$.}
\end{figure}

 As said before, in the $N=1$ case, one can simply take $Y_k=0$.
As seen in figure \ref{wkD2N1normafuera}, when the renormalization factor $Z_k$ 
in the regulator profile is chosen for values of
$\rho$ in the ``external'' part of the effective potential, the flow collapses 
after a certain renormalization-group ``time''. When $\rho_{text ren}>\rho_0$,
the flow blows at a finite RG time because there is no barrier preventing the singularity to be reached. The reasons
are the following, i) In the external part, the function $Z_k(\rho)$ rapidly 
stabilizes and accordingly the function $Z_k$ becomes a constant below a finite value of $k$, ii) The flow of $Z_k(\rho)$ in the 
``internal'' part continues to grow without bound, iii) For all $k>0$ in the 
internal part the potential is not
convex. Accordingly, the regulator becomes negligible in the internal
part of the potential and the flattening of the internal part is not strong 
enough to avoid the singularity (as is the case for the LPA with an 
exponential
regulator with $\alpha<2$). For $N=1$ we therefore choose $Z_k$ as 
the value of $Z_k(\rho)$ at $\rho=0$.

\begin{figure}[ht]
\centering\includegraphics[width=8cm]{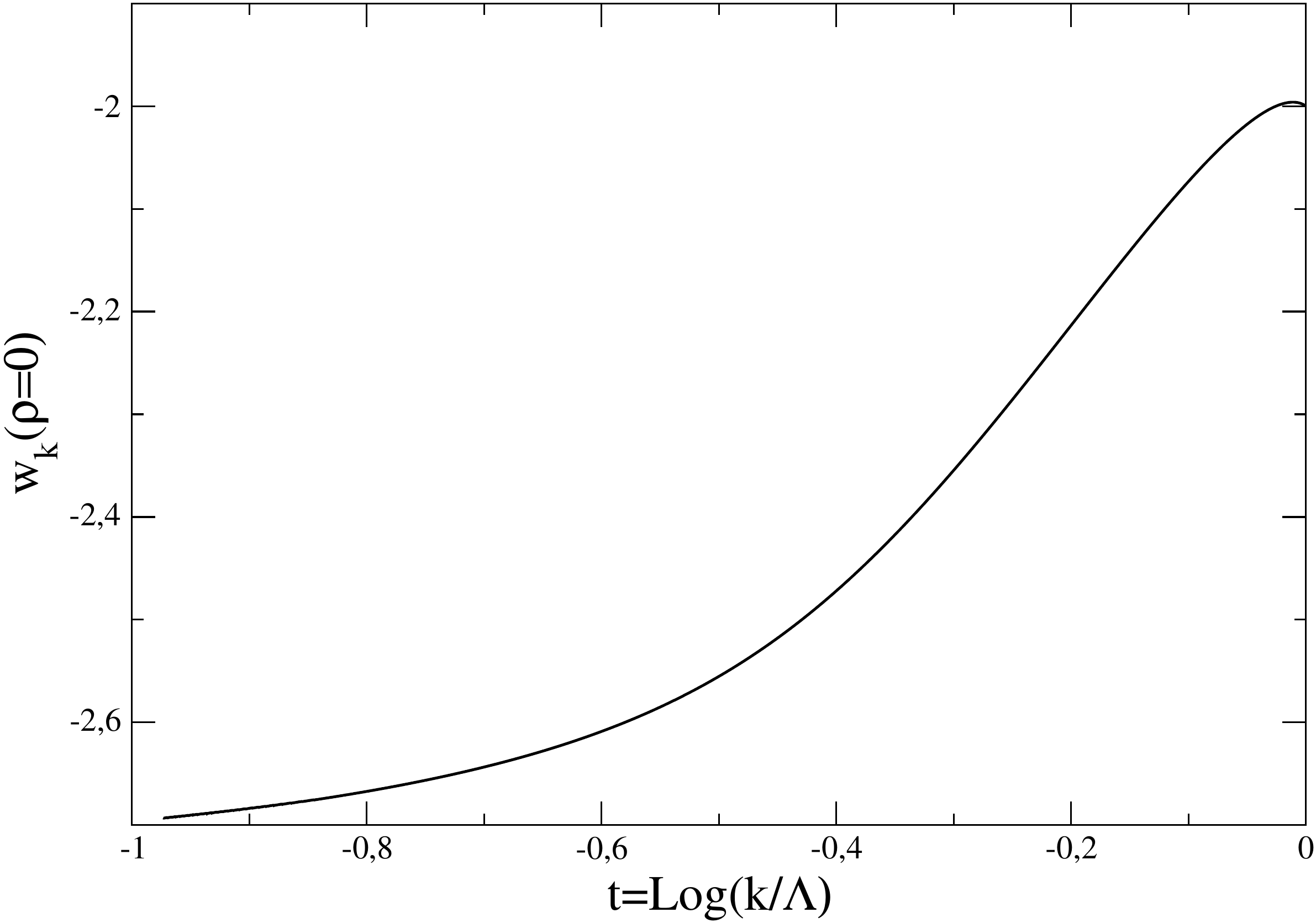}\\
\vspace{1cm}
\centering\includegraphics[width=8cm]{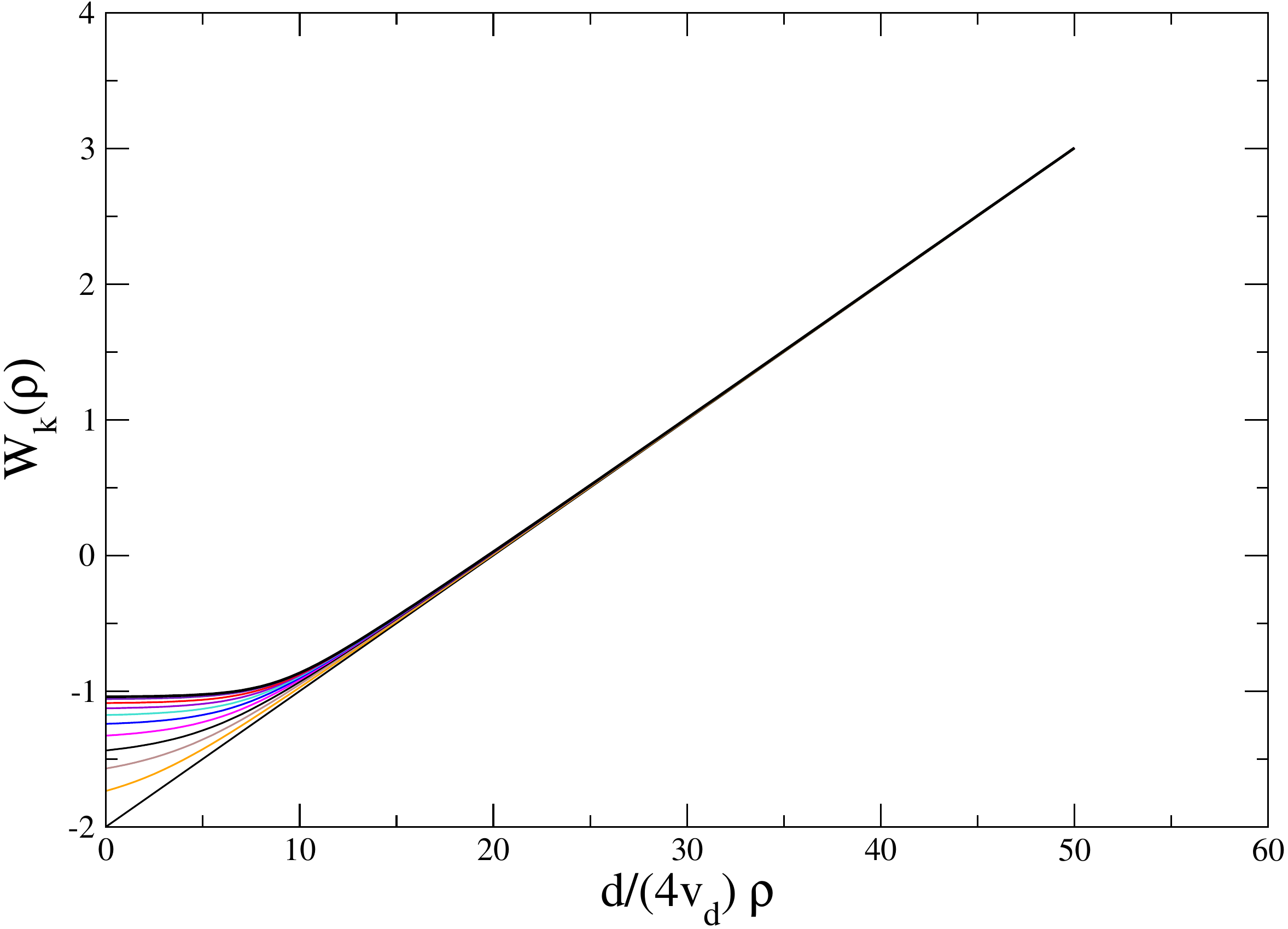} 
\caption{\label{D2N1d3} (Top) $w_k(\rho=0)$ as a function of 
$t=\log(k/\Lambda)$. (Bottom, color online) $W_k(\rho)$ as a function of $\rho$ 
for various values of $t$.
In both figures, $N=1$ and $d=3$ using the exponential regulator (\ref{expreg}) with 
$\alpha=3$ and normalization fixed at $\rho=0$. }
\end{figure}

\begin{figure}[ht]
\centering\includegraphics[width=8cm]{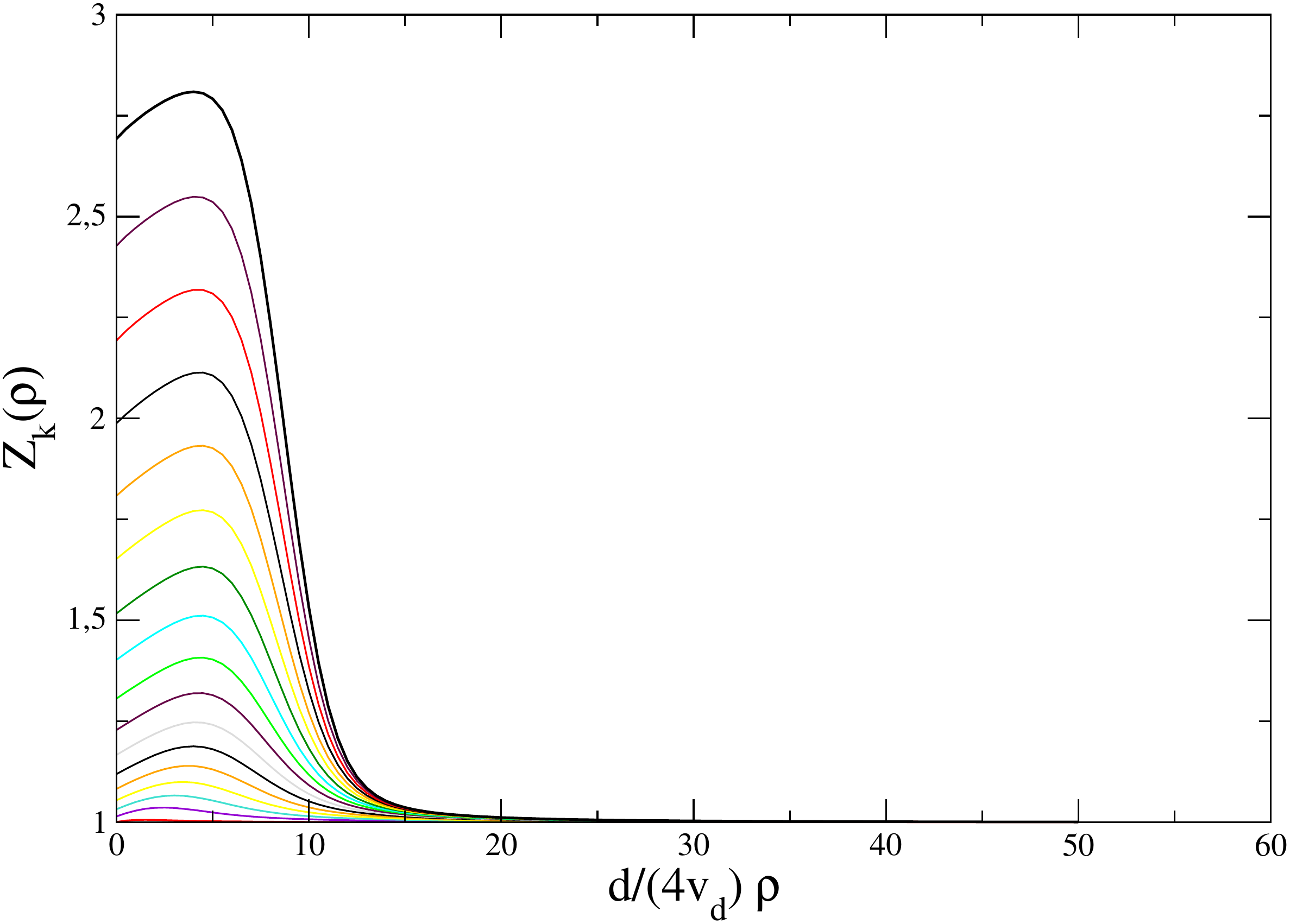} 
\caption{\label{ZkD2N1norma0} (color online) $Z_k(\rho)$ for $N=1$ and $d=3$ 
using the exponential regulator (\ref{expreg})
with $\alpha=3$. Normalization fixed at $\rho=0$.}
\end{figure}
\begin{figure}[ht]
\centering\includegraphics[width=8cm]{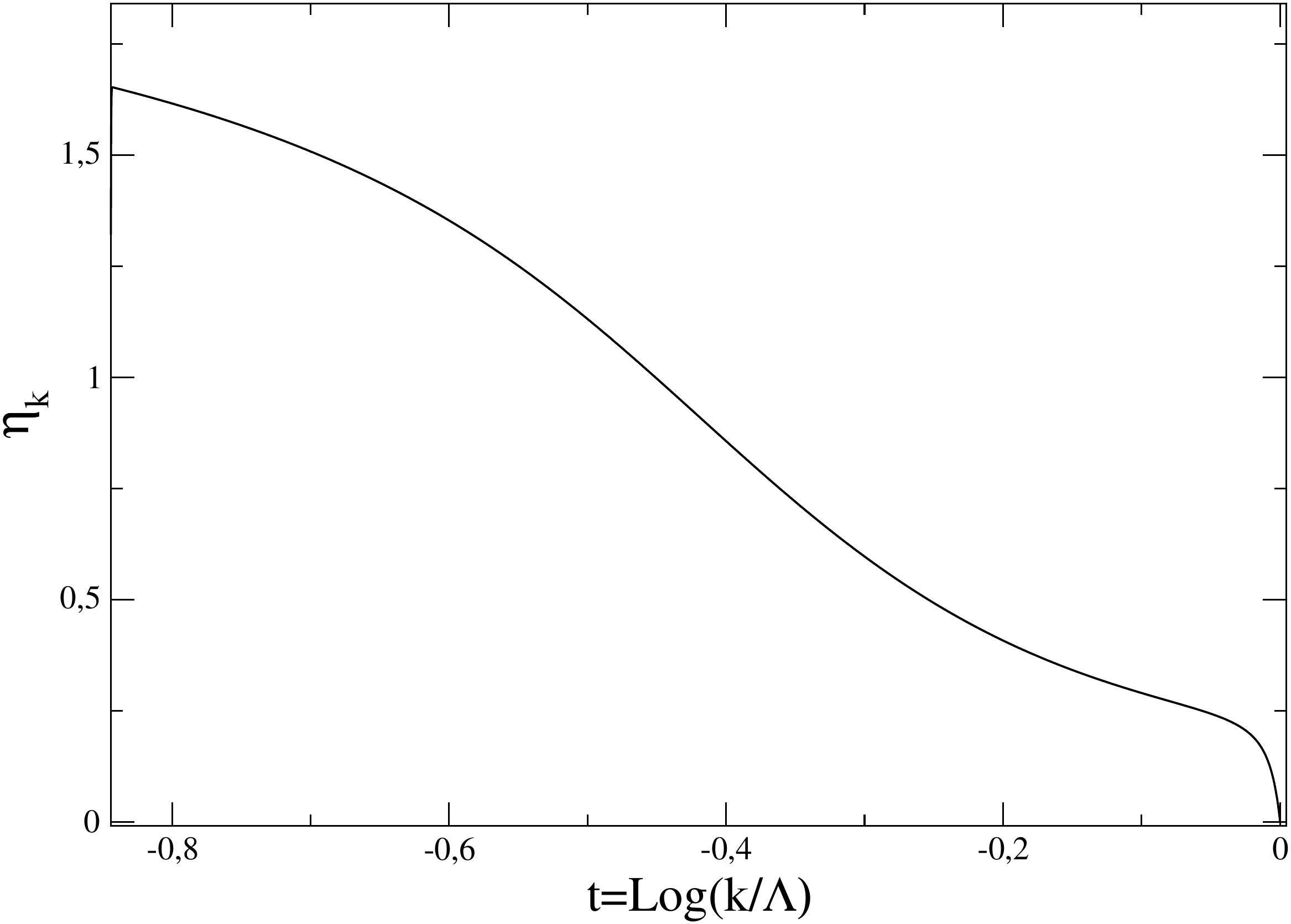} 
\caption{\label{etaD2N1norma0} $\eta_k$ for $N=1$ and $d=3$ using the 
exponential regulator (\ref{expreg})
with $\alpha=3$. Normalization fixed at $\rho=0$. Observe that, given that $Z_k(\rho=0)$
and $Z_k(\rho=2\rho_0)$ are different, the position of the singularity is different
from the LPA order.}
\end{figure}
\begin{figure}[ht]
\centering\includegraphics[width=8cm]{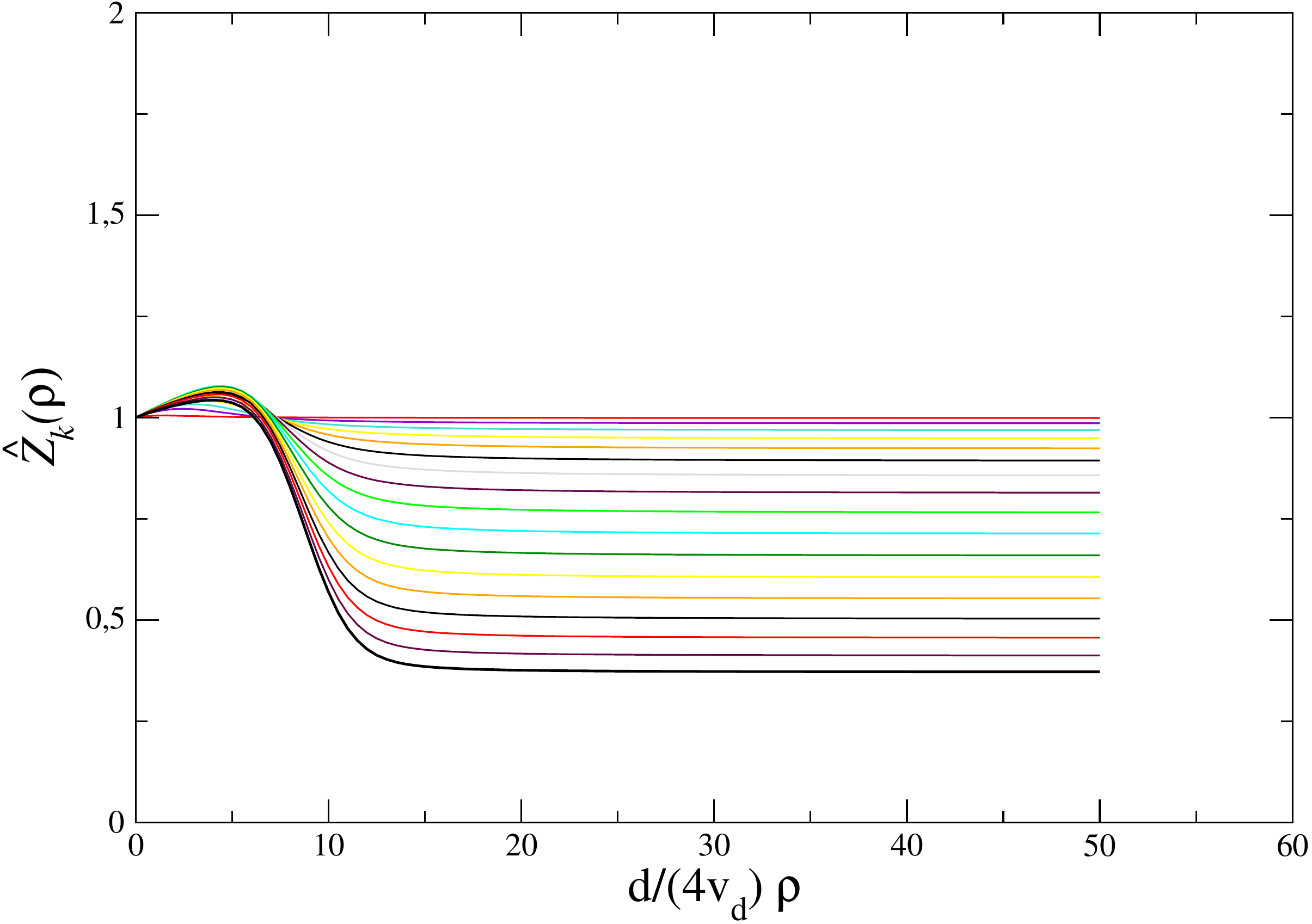} 
\caption{\label{ZknormD2N1norma0} (color online) $\hat Z_k(\rho)$ as function of 
$\rho$ for various values of $t$ for $N=1$ and $d=3$ using the exponential 
regulator (\ref{expreg})
with $\alpha=3$. Normalization fixed at $\rho=0$.}
\end{figure}

With this choice, we observe as in the LPA case that the effective
potential runs
in the low temperature phase to a convex potential that is flat in the 
``internal'' part but that finally collapses when, due to numerical lack of
precision, the singularity of the flow equation is crossed. We observe
systematically that if the discretization parameters of the program are improved, 
the instability
appears at larger renormalization-group ``times'' (indicating that this 
phenomenon is a numerical artifact), but, in practice, they are finally reached.
A typical run of the effective potential and of the function $Z_k(\rho)$
is shown in Figs.~\ref{D2N1d3} and \ref{ZkD2N1norma0} for $d=3$. In all cases,
we observe, on the top of a potential approaching the convexity, a function 
$Z_k(\rho)$ that seems to diverge when $k\to 0$ for values of $\rho$ in the 
``internal''
part of the potential. In order to study how this divergence takes place, we plot 
in Fig.~\ref{etaD2N1norma0} the quantity $\eta_k=-\partial_t \log Z_k$ that shows the
exponent of the divergence of $Z_k(\rho=0)$ as a function of $t$. 
We observe that the exponent seems to stabilize at values $\sim 1.5$ but at that 
value of $t$ the singularity is hit and the flow breaks down. In
Fig.~\ref{ZknormD2N1norma0} the renormalized function
\begin{equation}
 \hat Z_k(\rho)= \frac{Z_k(\rho)}{Z_k(\rho=0)}
\end{equation}
is plotted. It seems to approach a finite limit for values of $\rho$ 
corresponding to the ``internal'' part of the potential and, as expected, to 
tend to zero
in the ``external'' part (given the fact that the function $Z_k(\rho)$ seems to 
go to a finite limit in that regime). As a conclusion, the second order of
the DE seems to respect be able to mantain the convexity property and the
singularity present in the flow equation for the potencial does not seem to be hit.
However, a standard numerical implementation finally breaks down (as in the LPA)
because of the unavoidable lack of precision.

\begin{figure}[ht]
\centering\includegraphics[width=8cm]{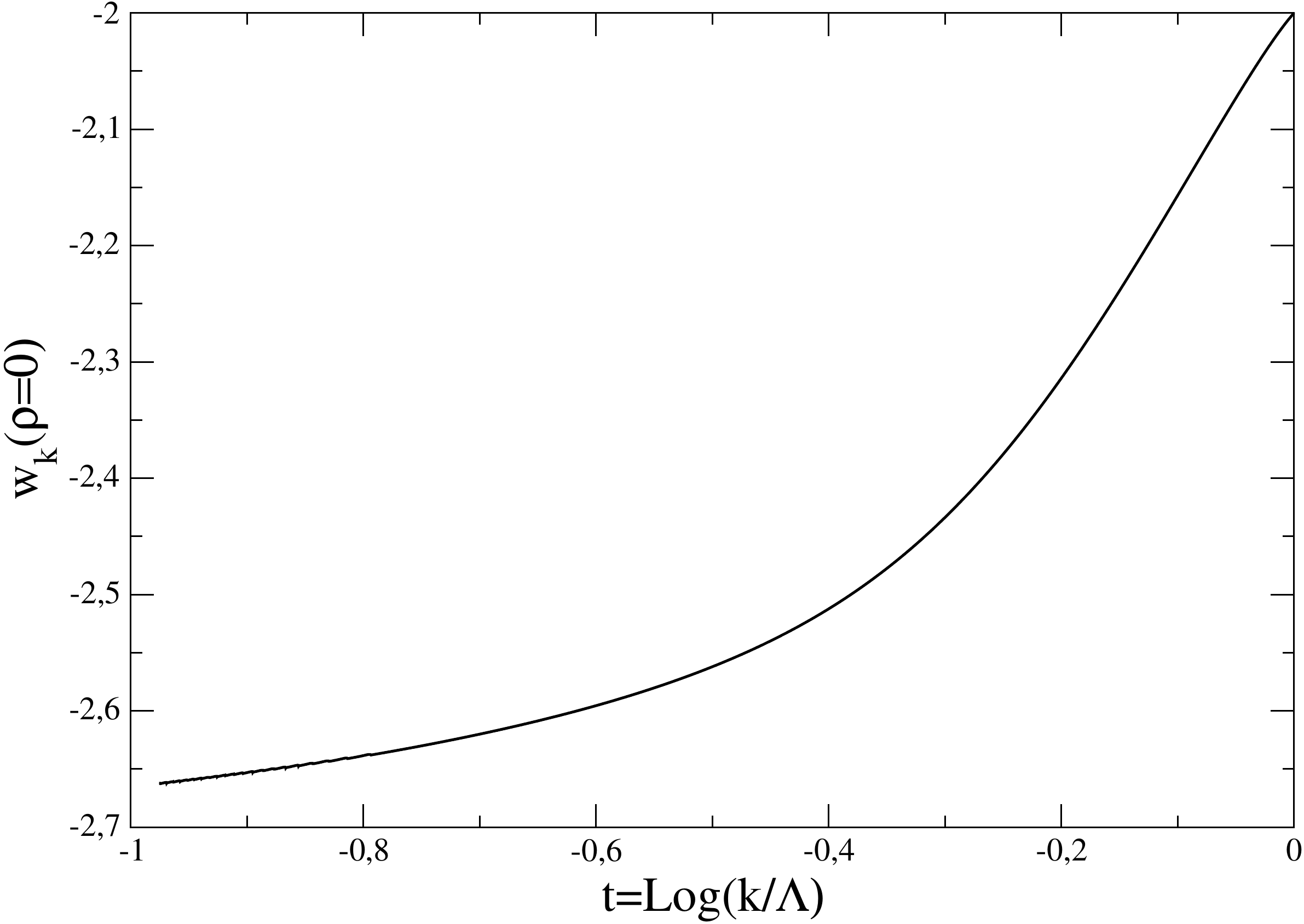}\\
\vspace{1cm}
\centering\includegraphics[width=8cm]{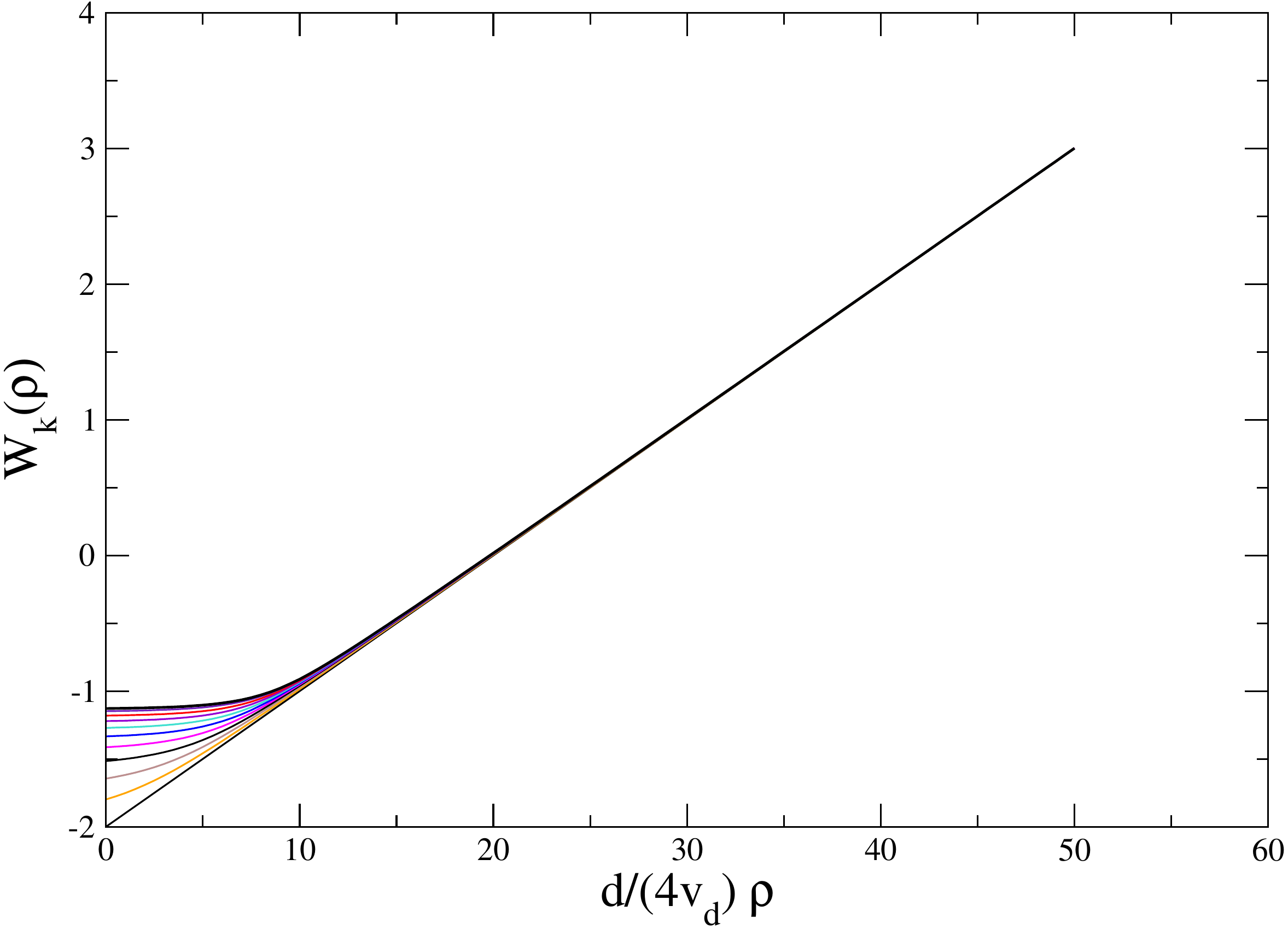} 
\caption{\label{D2N1d2potencial} (Top) $w_k(\rho=0)$ as a function of 
$t=\log(k/\Lambda)$. (Bottom, color online) $W_k(\rho)$ as a function of $\rho$ 
for different
values of $t$. In both figures, for $N=1$ and $d=2$ using the exponential 
regulator (\ref{expreg}) with $\alpha=3$ and normalization fixed at $\rho=0$.}
\end{figure}

\begin{figure}[ht]
\centering\includegraphics[width=8cm]{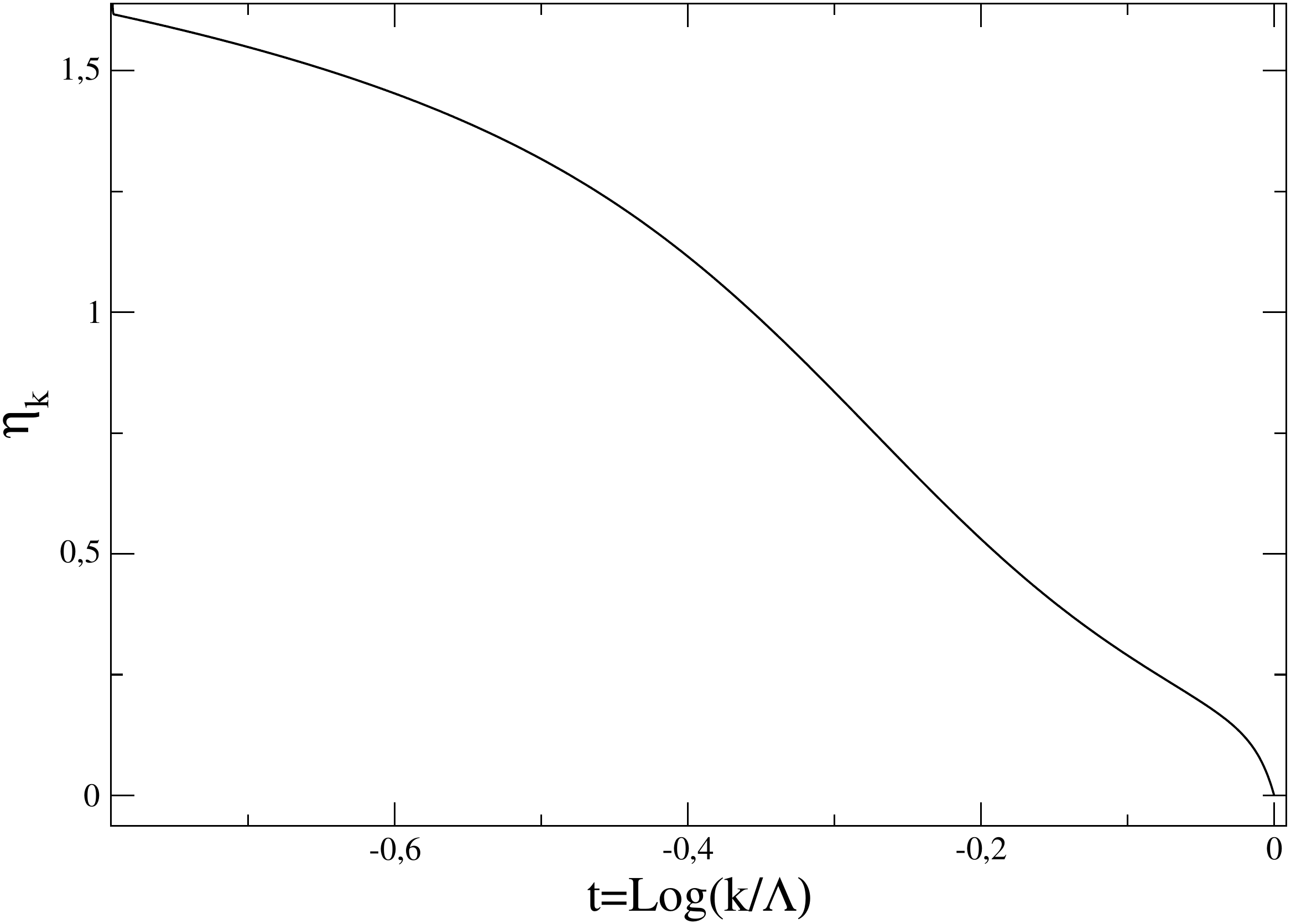}
\caption{\label{D2N1d2eta} $\eta_k$ for $N=1$ and $d=2$ using the exponential 
regulator (\ref{expreg})
with $\alpha=3$. Normalization fixed at $\rho=0$.}
\end{figure}

\begin{figure}
\includegraphics[width=8cm]{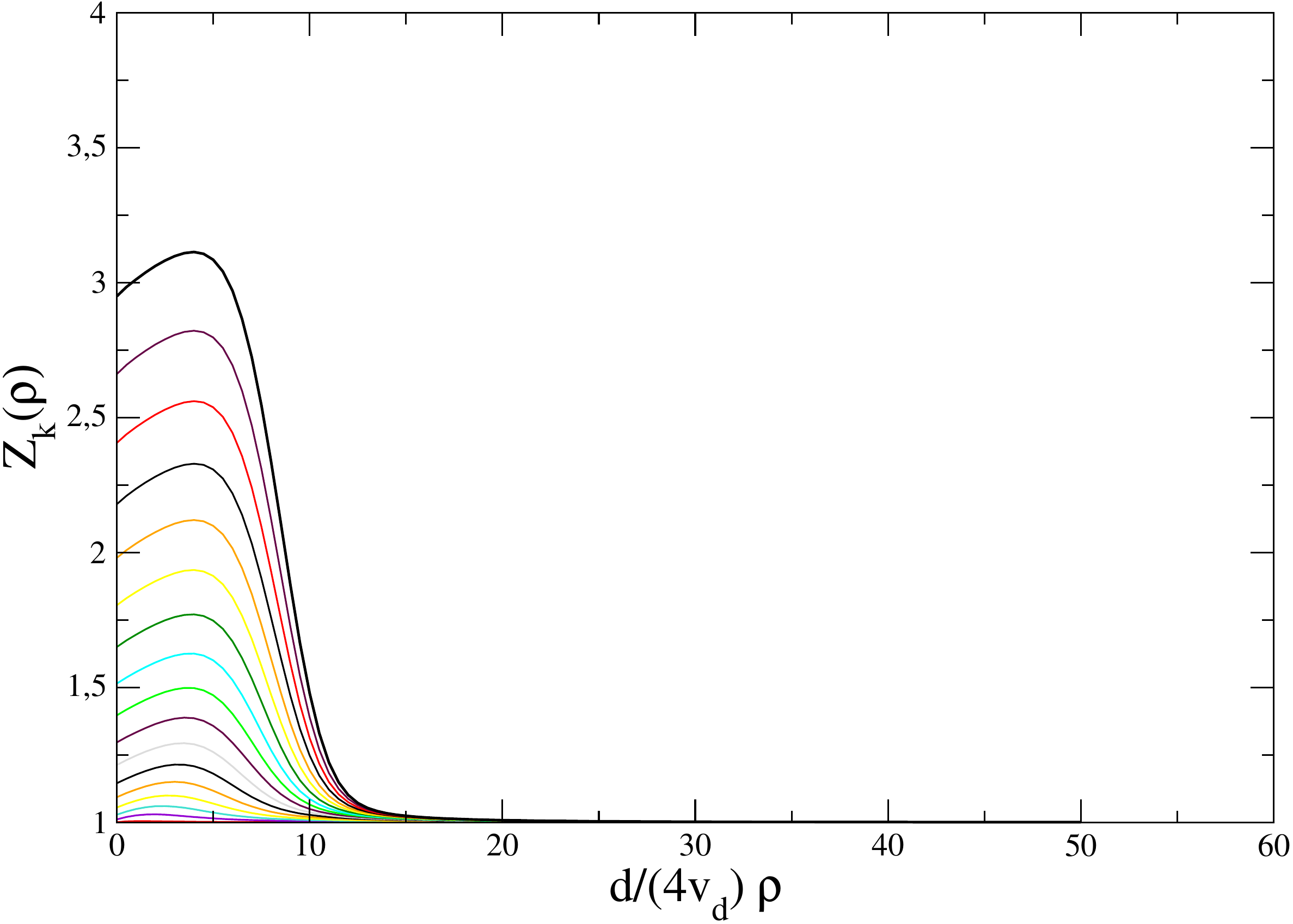}\\
\vspace{1cm}
\centering\includegraphics[width=8cm]{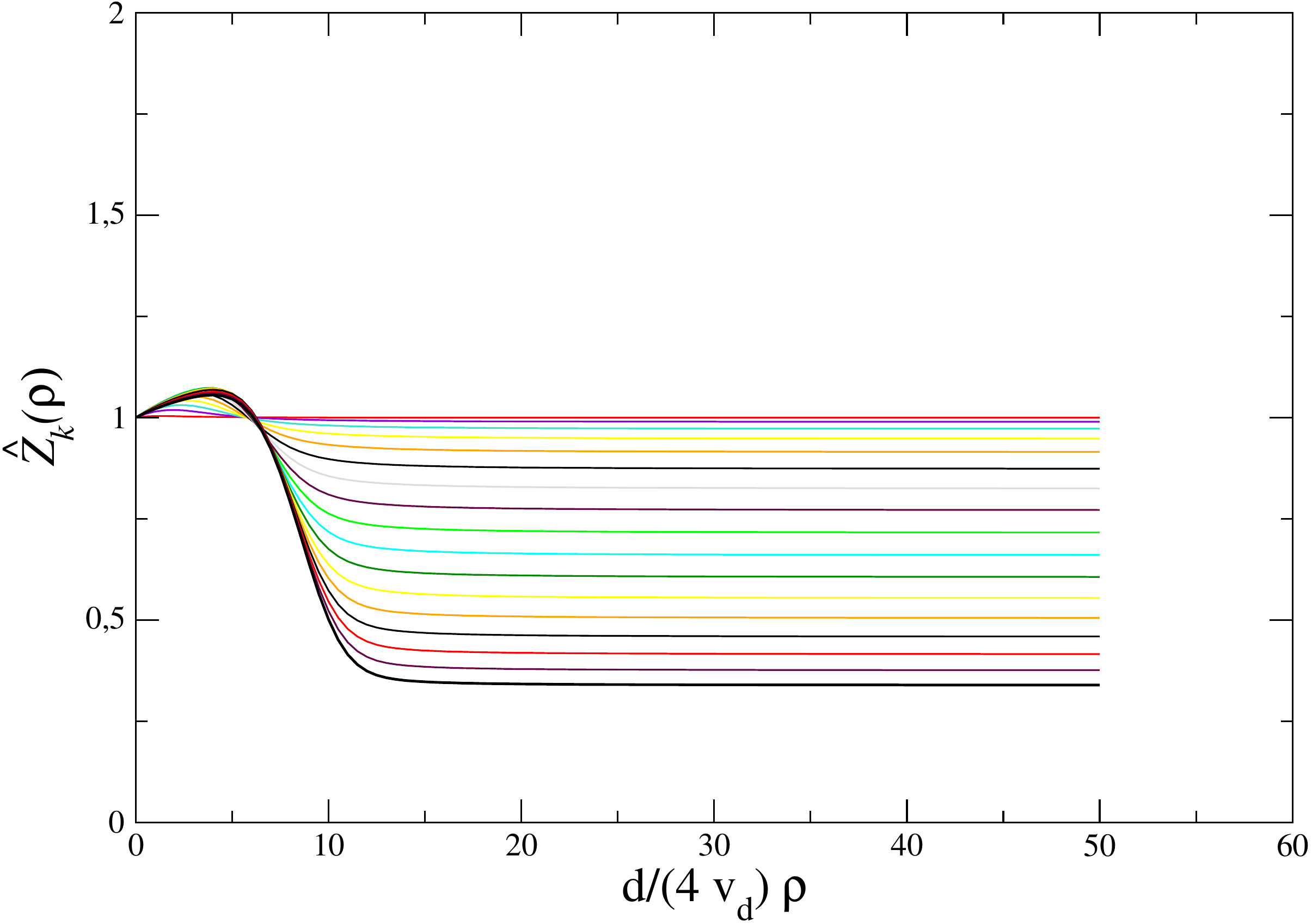} 
\caption{\label{D2N1d2} (Top, color online) $Z_k(\rho)$ as a function of $\rho$. 
(Bottom, color online) $\hat Z_k(\rho)$
as function of $\rho$. In both cases,  for various values of $t$ for $N=1$ and 
$d=2$ using the exponential regulator (\ref{expreg})
with $\alpha=3$. Normalization fixed at $\rho=0$.}
\end{figure}

As said before, contrarily to what
happens at the LPA level, the second order of the DE clearly shows for the 
single scalar case
a low temperature phase not only in $d=3$ but also in $d=2$. This is shown in Figs.~ 
\ref{D2N1d2potencial},\ref{D2N1d2eta} and \ref{D2N1d2}.
As explained before, this is one of the most important ingredients absent at the 
LPA level and one of the main reasons to go beyond. The results for $d=2$ seem to
be qualitatively similar to those of $d=3$ for $N=1$.


\subsection{Generic $O(N)$ model}

In this section we extend the analysis of the second order of the derivative 
expansion in the low temperature phase for $N>1$.

\begin{figure}
\centering\includegraphics[width=8cm]{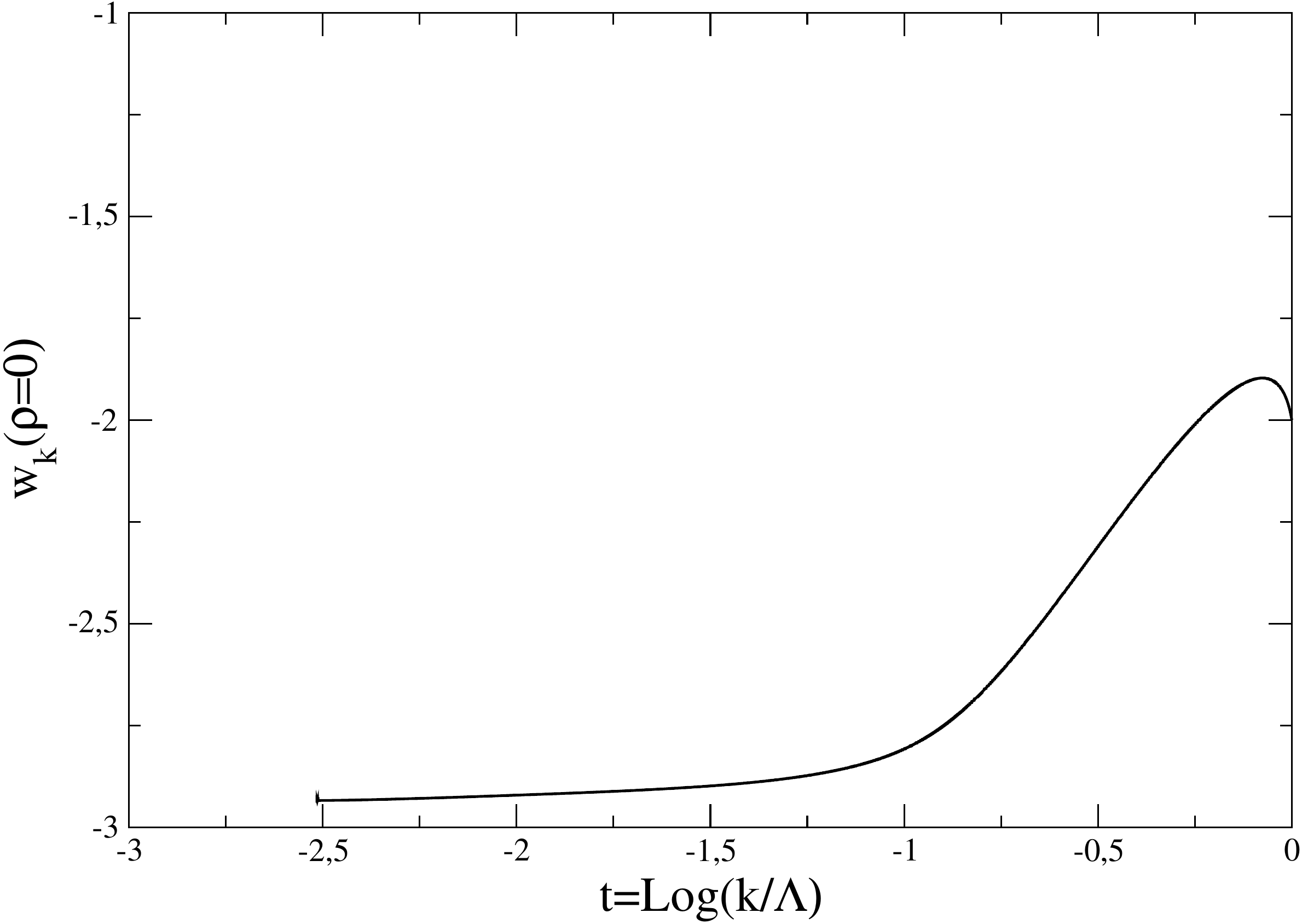}\\
\vspace{1cm}
\includegraphics[width=8cm]{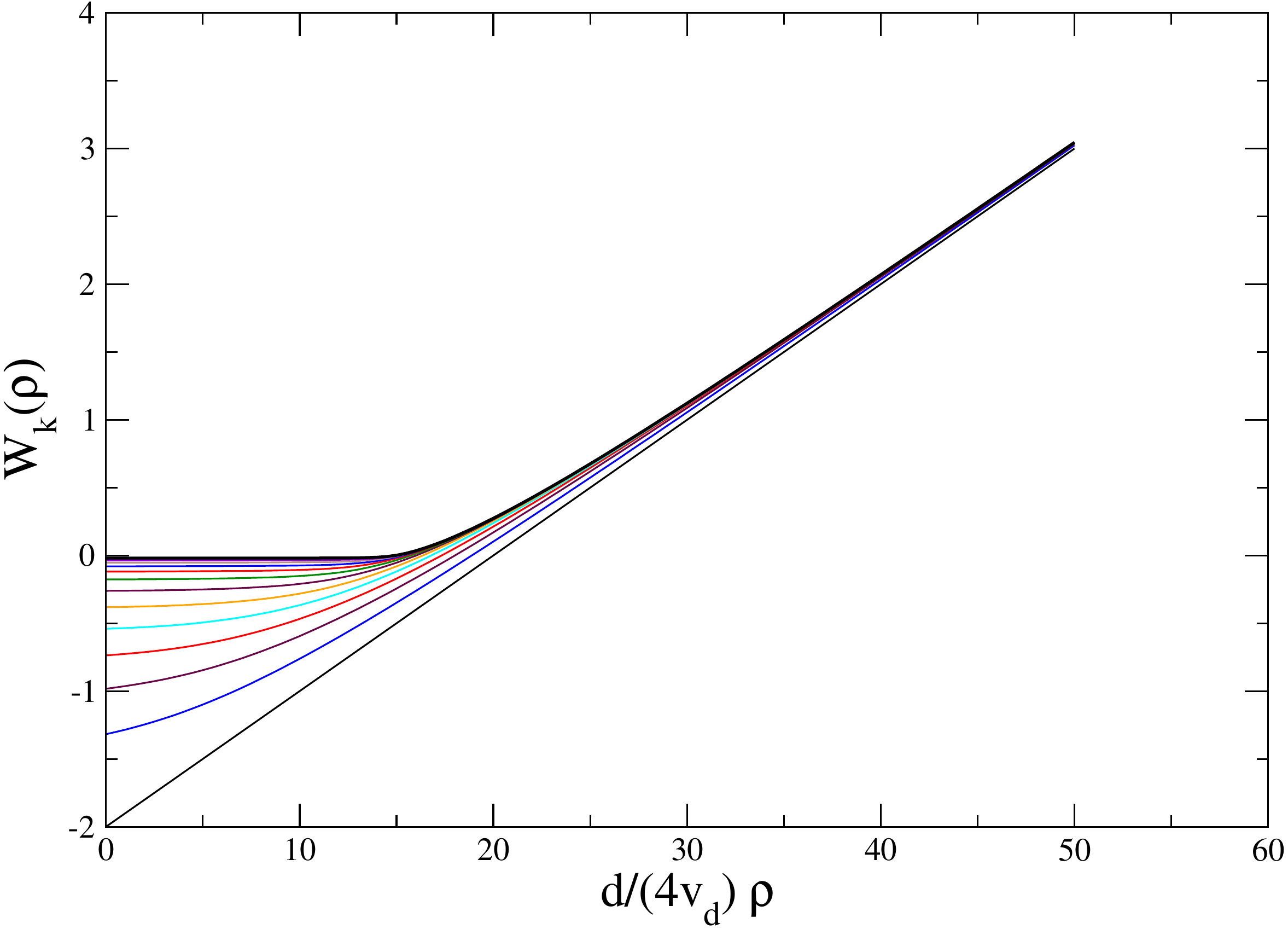} 
\caption{\label{D2N4d3} (Top) $w_k(\rho=0)$ as a function of 
$t=\log(k/\Lambda)$. (Bottom, color online) $W_k(\rho)$ as a function of $\rho$ 
for various values of $t$.
In both figures, for $N=4$ and $d=3$ using the exponential regulator (\ref{expreg}) with 
$\alpha=3$ and normalization fixed at $\rho=2\rho_0$.}
\end{figure}
\begin{figure}
\centering\includegraphics[width=8cm]{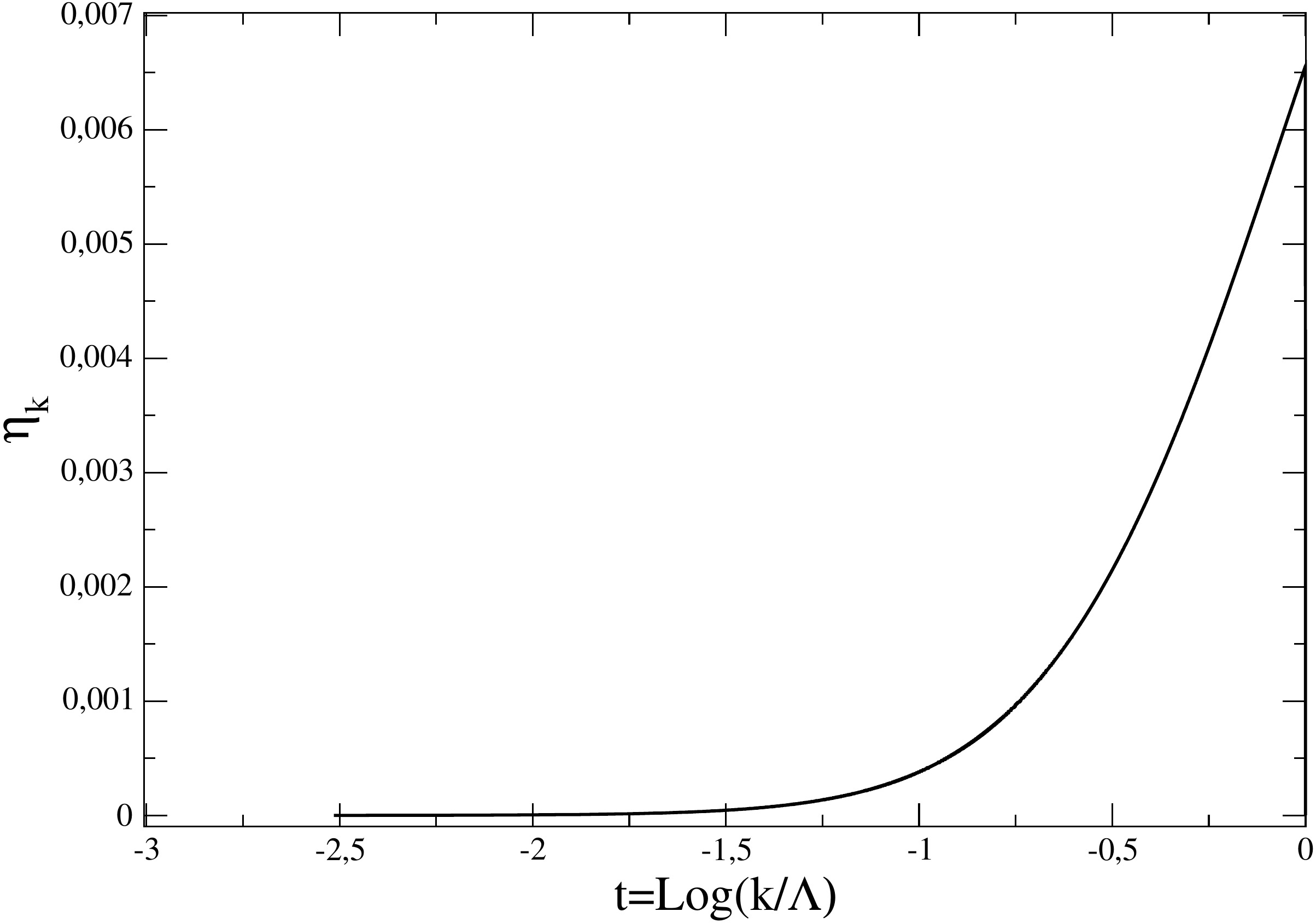} 
\caption{\label{D2N4d3biseta} $\eta_k$ as a function of $t$ for $N=4$ and $d=3$ 
using the exponential regulator (\ref{expreg})
with $\alpha=3$. Normalization fixed at $\rho=2\rho_0$.}
\end{figure}

In this case, in contrast to the $N=1$ case,
choosing $Z_k$ at values larger than $\rho_0$ turns out to give a more 
stable flow than the one obtained if $Z_k$ is fixed at $\rho$ smaller than $\rho_0$.
When $Z_k$ is fixed at values larger than $\rho_0$  the flow does not explode until large values of $|t|$.
Fig.\ref{D2N4d3} shows that, as for LPA, a convex potential is approached along the flow. As 
before, we need to choose a value of $\alpha$ larger than $2$ to avoid hitting the singularity.
We employed an Euler algorithm of the same kind that the one employed in the LPA 
(without the improvement in the internal part of the potential).
Even if convexity is clearly visible we are not able to reach very 
large values of $|t|$, because of the same numerical instabilities discussed 
before. In
any case, as for the LPA approximation, the flow for $N>1$ is much more stable than 
for the $N=1$.
It is very plausible that an hybrid algorithm that exploits
the exact behavior of NPRG equations in the internal region would, as for the LPA 
case, allow to make the flow even more stable.
\begin{figure}
\centering\includegraphics[width=8cm]{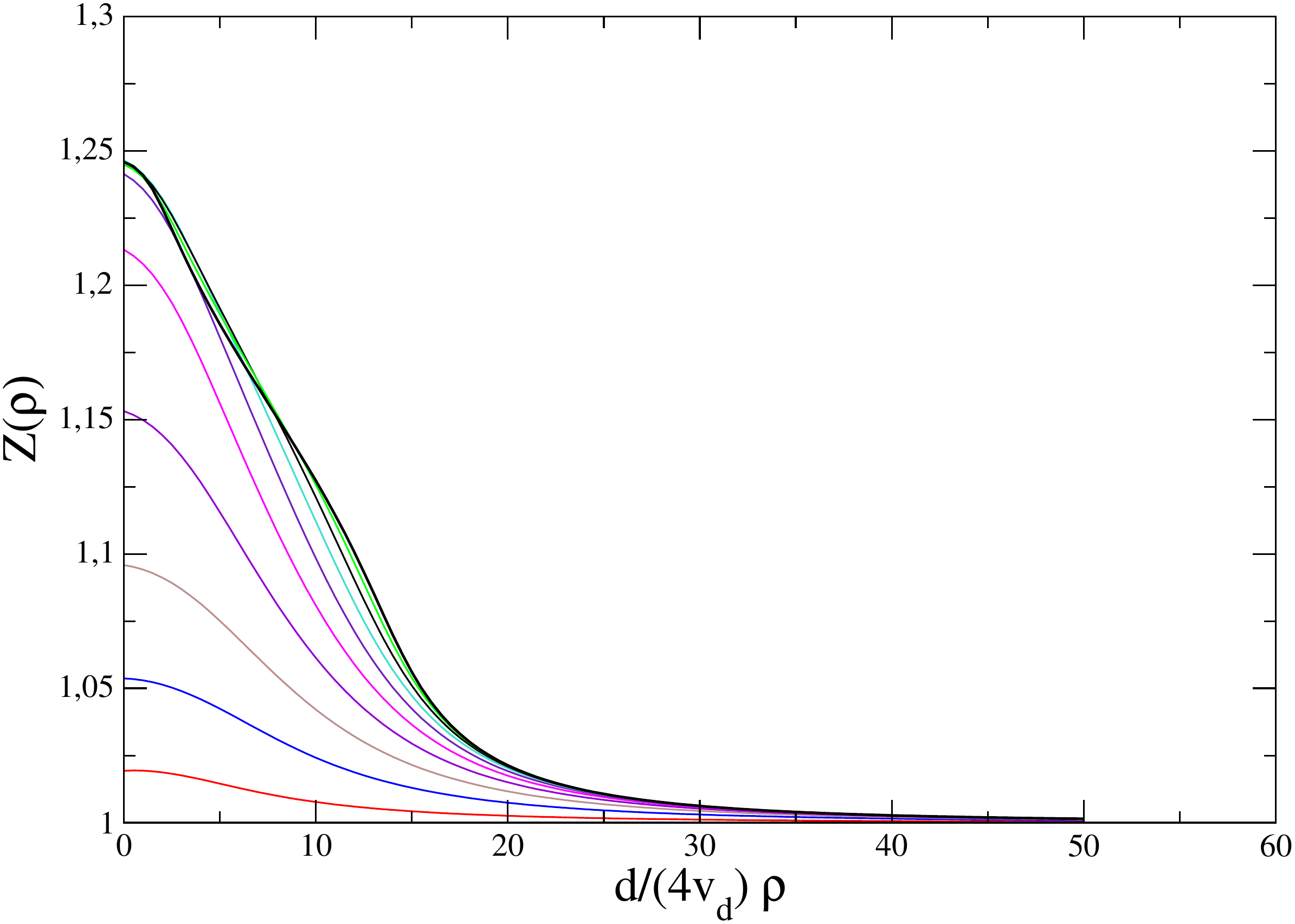}\\
\vspace{1cm}
\centering\includegraphics[width=8cm]{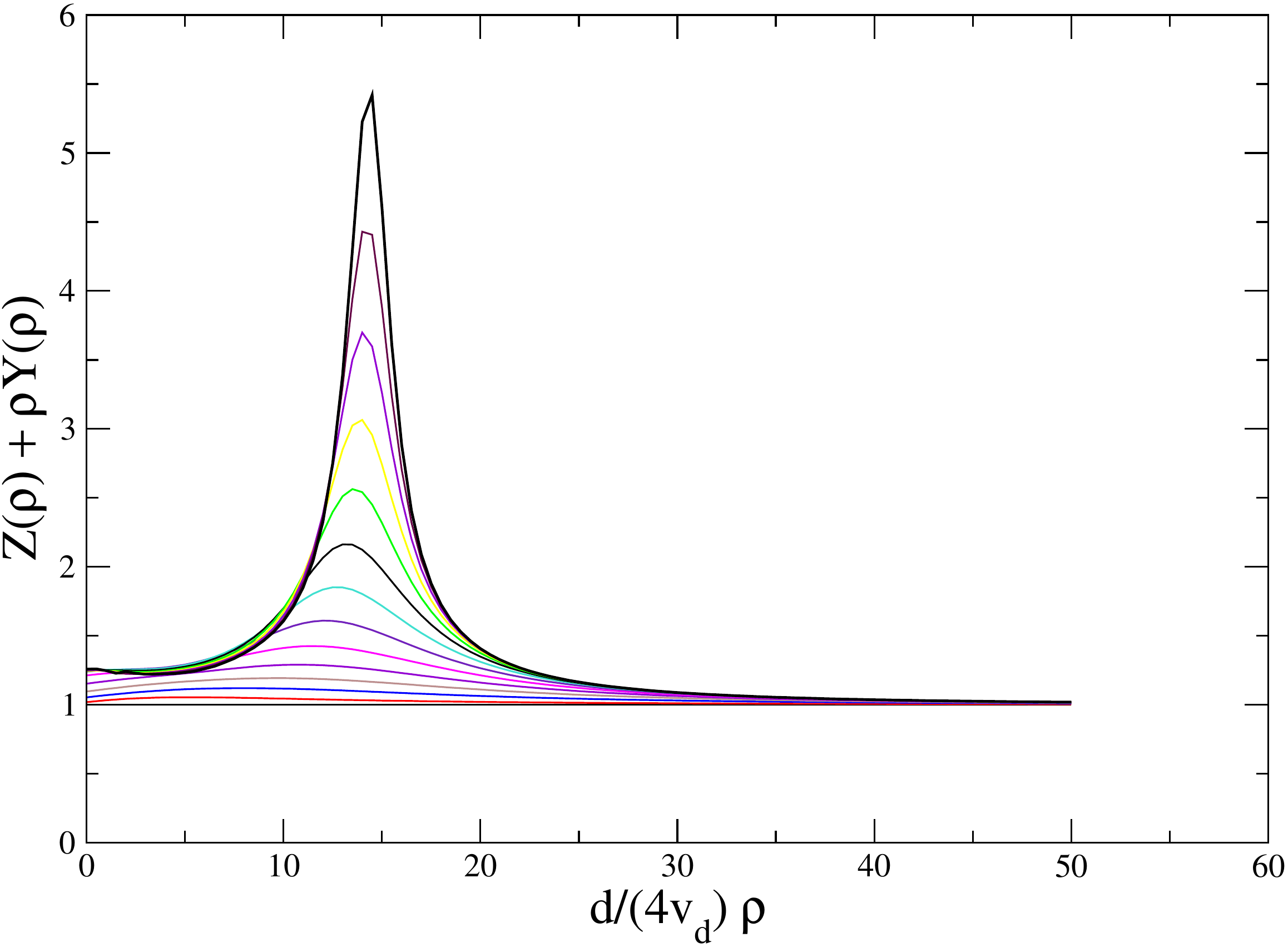}
\caption{\label{D2N4d3bis} (Top, color online) $Z_k(\rho)$ as a function of 
$\rho$. (Bottom, color online) $Z_k(\rho)+\rho Y_k(\rho)$
as function of $\rho$. In both cases,  for various values of $t$ for $N=4$ and 
$d=3$ using the exponential regulator (\ref{expreg})
with $\alpha=3$. Normalization fixed at $\rho=2\rho_0$.}
\end{figure}

It is interesting to discuss also the results for the flows of the functions
$Z_k(\rho)$ and $Y_k(\rho)$. In fact, we prefer to present the results in terms 
of the transverse renormalization function $Z_k(\rho)$ and the longitudinal one,
$Z_k(\rho)+\rho Y_k(\rho)$, as shown in Figs.~\ref{D2N4d3biseta} and \ref{D2N4d3bis}. As one can observe, 
the function $Z_k(\rho)$ seems to reach a limit when $k\to 0$. In contrast,
the longitudinal renormalization
factor $Z_k(\rho)+\rho Y_k(\rho)$ seems to grow without bound around the minimum of the potential. This is 
similar to what is observed in $d=4$ in \cite{Zappala12}. However, in the
$d=3$ the effect is much stronger. In what concerns the behavior of these 
functions in the ``external'' region of the potential, it seems to stabilize 
faster.
This can be seen in the running of the anomalous dimension (fixed via the value of 
$Z_k$ at $2\rho_0$) as can be seen in Fig.~\ref{D2N4d3biseta}.
One observes that $\eta_k$ goes to zero as expected
in the low temperature phase.

We have thus managed to study the low temperature phase of $O(N)$ models in the second order of the
DE and shown that if an appropriate regulator and renormalization condition is used, one can show clearly
that the property of convexity of the effective potential is respected. The flow finally becomes unstable
for numerical reasons. When $N>1$ the flow is much more stable than in the single scalar case $N=1$.

\section{Conclusions}

In the present article we analyze the NPRG equations both at the leading order 
(LPA) of the Derivative Expansion and at next-to-leading order (order 
$\mathcal{\partial}^2$) in
the low temperature phase. These simple approximations performed at the level of 
NPRG equations are able, in contrast with most perturbative schemes, to preserve the 
convexity
of the free energy. In the present article we show that this is only true for certain
regulators. In particular, the most used 
regulators used (the $\theta$-regulator and the exponential regulator) are able 
to respect the convexity of the free energy. However, in the case
of the exponential regulator, it is necessary to choose a  large 
enough pre-factor. If this is not done, a singularity of the flow is hit.

Even if an appropriate regulator is chosen, there is a practical difficulty:
the flow approaches a singularity without crossing it. As a 
consequence, even if the singularity is never hit, the flow 
becomes numerically unstable
for low enough values of $k$. In order to deal with this problem many 
algorithms have been proposed in the literature. We implement at the LPA level a very 
simple algorithm that
exploits the exact behavior of the flow in the ``internal'' part of the 
potential and that makes the flow stable for essentially arbitrary values of 
$k$ when $N>1$. The
$N=1$ case turns out to be much more challenging for various reasons discussed along 
the article. The most important one comes from the fact that the physical 
longitudinal susceptibility is
a continuous function of the external field for $N>1$ but has a discontinuity for
$N=1$ when $k\to 0$. This makes this case much harder to treat. We obtain for $N>1$ and with a very simple algorithm
a flow that is 
qualitatively more stable than what was obtained before.

On top of this analysis, we studied the behavior of the flow in the low 
temperature phase at order $\mathcal{\partial}^2$ of the Derivative Expansion. 
We observe that
in order to approach a convex free-energy it is necessary to normalize the field 
in a different way in the case $N=1$ and in the case $N>1$. On one hand, in 
order to avoid reaching the
singularity of the flow for $N=1$ it is necessary to normalize the field in the 
``internal'' region of the potential. On the other hand, in the $N>1$ case it is 
necessary
to normalize the field in the ``external'' part of the potential.
Once these 
choices are made and an appropriate regulator is chosen, the flow 
approaches
a convex free-energy. Of course, in practice, the flow becomes 
numerically unstable when $k$ is very small and the singularity is 
reached in practice.\footnote{The two-dimensional XY case \cite{Kosterlitz:1973xp} has been studied in references \cite{VonGersdorff:2000kp,Jakubczyk:2014isa}.
In that
case, the singularity seems always reached \cite{Jakubczyk:2014isa}. It must be stressed that this low temperature phase is very peculiar.}

For the future, we are planing to implement the same kind of algorithm that we 
presented in the LPA case at second order of the Derivative Expansion 
$\mathcal{O}(\partial^2)$.
We are planing also to make the same kind of analysis in more elaborated 
approximations such as the one proposed in \cite{Blaizot05,Benitez09,Benitez12}. We 
would like also to try to implement an improved
algorithm in these kinds of approximations also. These improved approximations 
and algorithms in the low temperature phase of $O(N)$ models could be useful in 
the analysis of
a large variety of physical problems, that we are planing to analyze,
 as the formation of bound states or
the calculation of the phase diagram for realistic microscopic models.

\begin{acknowledgments}
 The authors acknowledge financial support from the ECOS-Sud France-Uruguay 
program U11E01, and from the PEDECIBA.
 We  thank also  B. Delamotte, N. Dupuis and M. Tissier for useful discussions.
\end{acknowledgments}

\bibliographystyle{unsrt}

\begin{thebibliography}{10}

\bibitem{Wetterich93}  C.Wetterich, Phys. Lett., {\bf B301}, 90 (1993).

\bibitem{Ellwanger93}  U.Ellwanger, Z.Phys., {\bf C58}, 619 (1993).

\bibitem{Tetradis94}
  N.~Tetradis and C.~Wetterich,
  Nucl.\ Phys.\ B {\bf 422},  541 (1994).


\bibitem{Morris94}  T.R.Morris, Int. J. Mod. Phys., {\bf A9}, 2411 (1994).

 \bibitem{Morris94c}
T.~R. Morris,
Phys. Lett. B329 (1994) 241--248.


\bibitem{Berges02}
J. Berges, N. Tetradis and C. Wetterich,
   Phys. Rept. {\bf 363}, 223--386 (2002).
   
\bibitem{Delamotte07} 
  B.~Delamotte,
  cond-mat/0702365 [COND-MAT].
   
\bibitem{Ringwald89} 
  A.~Ringwald and C.~Wetterich,
  Nucl.\ Phys.\ B {\bf 334}, 506 (1990).
   
\bibitem{Tetradis92} 
  N.~Tetradis and C.~Wetterich,
theories,''
  Nucl.\ Phys.\ B {\bf 383}, 197 (1992).
  
\bibitem{Horikoshi98} 
  A.~Horikoshi, K.~-I.~Aoki, M.~-A.~Taniguchi and H.~Terao,
  hep-th/9812050.
  
\bibitem{Alexandre98} 
  J.~Alexandre, V.~Branchina and J.~Polonyi,
  Phys.\ Lett.\ B {\bf 445}, 351 (1999)
  [cond-mat/9803007].
  
\bibitem{Kapoyannis00} 
  A.~S.~Kapoyannis and N.~Tetradis,
  Phys.\ Lett.\ A {\bf 276}, 225 (2000)
  [hep-th/0010180].
  
\bibitem{Caillol12} 
  J.~-M.~Caillol,
  Nucl.\ Phys.\ B {\bf 855}, 854 (2012).
  
\bibitem{Zappala12} 
  D.~Zappala,
  Phys.\ Rev.\ D {\bf 86}, 125003 (2012)
  [arXiv:1206.2480 [hep-th]].
  
\bibitem{Canet11} 
  L.~Canet, H.~Chat\'e and B.~Delamotte,
  J.\ Phys.\ A A {\bf 44}, 495001 (2011)
  [arXiv:1106.4129 [cond-mat.stat-mech]].

  \bibitem{Tissier11} 
  M.~Tissier and G.~Tarjus, Phys.\ Rev.\ Lett.\ {\bf 107}, 041601 (2011).

    \bibitem{Golner86} 
  G.~R.~ Golner, Phys. Rev. {\bf B33}, (1986) 7863.

  \bibitem{Canet03}
L. Canet, B. Delamotte, D. Mouhanna and J. Vidal,
 Phys. Rev. B68 (2003) 064421.

\bibitem{Chate12prox}
L.~Canet, H.~Chat\'e and B.~Delamotte, unpublished.

\bibitem{Pangon09} 
  V.~Pangon, S.~Nagy, J.~Polonyi and K.~Sailer,
  Int.\ J.\ Mod.\ Phys.\ A {\bf 26}, 1327 (2011)
  [arXiv:0907.0144 [hep-th]].
  
\bibitem{Pangon10} 
  V.~Pangon,
  Int.\ J.\ Mod.\ Phys.\ A {\bf 227}, 1250014 (2012)
  [arXiv:1008.0281 [hep-th]].

\bibitem{Blaizot05}
  J.-P.~Blaizot, R.~M\'endez Galain and N.~Wschebor,
  Phys.\ Lett.\ B {\bf 632}, 571 (2006)

\bibitem{Benitez09} 
  F.~Benitez, J.~-P.~Blaizot, H.~Chat\'e, B.~Delamotte, R.~M\'endez-Galain and 
N.~Wschebor,
  Phys.\ Rev.\ E {\bf 80}, 030103 (2009)
  [arXiv:0901.0128 [cond-mat.stat-mech]].
  
\bibitem{Benitez12} 
  F.~Benitez, J.-P.~Blaizot, H.~Chat\'e, B.~Delamotte, R.~Mendez-Galain and 
N.~Wschebor,
  Phys.\ Rev.\ E {\bf 85}, 026707 (2012)
  [arXiv:1110.2665 [cond-mat.stat-mech]].

\bibitem{Ellwanger94a}
U. Ellwanger,
 Z. Phys. C{\bf 62}  503--510 (1994).
 
 \bibitem{Morris94b}
T.~R. Morris,
Int. J. Mod. Phys. A9 2411--2450 (1994).  

\bibitem{Machado10} 
  T.~Machado and N.~Dupuis,
  Phys.\ Rev.\ E {\bf 82}, 041128 (2010)
  [arXiv:1004.3651 [cond-mat.stat-mech]].
  
\bibitem{Rancon11} 
  A.~Rancon and N.~Dupuis,
  Phys.\ Rev.\ B {\bf 84}, 174513 (2011)
  [arXiv:1106.5585 [cond-mat.quant-gas]].

\bibitem{Litim}
D.~Litim, Phys. Lett. {\bf B486}, 92 (2000); Phys. Rev. {\bf D64},
105007 (2001);  Nucl. Phys. {\bf B631}, 128 (2002); Int.J.Mod.Phys.
{\bf A16}, 2081 (2001).  

\bibitem{Canet02b} 
  L.~Canet, B.~Delamotte, D.~Mouhanna and J.~Vidal,
  Phys.\ Rev.\ D {\bf 67}, 065004 (2003)
  [hep-th/0211055].

\bibitem{Tetradis95}
N.~Tetradis and D.~F. Litim,
Nucl. Phys.  B464 (1996) 492--511.  

\bibitem{D'Attanasio97} 
  M.~D'Attanasio and T.~R.~Morris,
  Phys.\ Lett.\ B {\bf 409}, 363 (1997)
  [hep-th/9704094].


\bibitem{Kosterlitz:1973xp} 
  J.~M.~Kosterlitz and D.~J.~Thouless,
  J.\ Phys.\ C {\bf 6}, 1181 (1973).
  
\bibitem{Delamotte:2015aaa} 
  B.~Delamotte, M.~Tissier and N.~Wschebor,
  arXiv:1501.01776 [cond-mat.stat-mech].

\bibitem{VonGersdorff:2000kp} 
  G.~Von Gersdorff and C.~Wetterich,
  Phys.\ Rev.\ B {\bf 64}, 054513 (2001)
  [hep-th/0008114].
  
  
\bibitem{Jakubczyk:2014isa} 
  P.~Jakubczyk, N.~Dupuis and B.~Delamotte,
  Phys.\ Rev.\ E {\bf 90}, no. 6, 062105 (2014)
  [arXiv:1409.1374 [cond-mat.stat-mech]].
  
  
\end{thebibliography}

\end{document}